\documentclass[10pt]{article}
\usepackage{amsthm,amsmath,amssymb,color}
\usepackage[hyphens]{url}
\usepackage[utf8]{inputenc}
\usepackage[margin=2cm]{geometry}
\usepackage[normalem]{ulem}
\usepackage{amssymb,amsmath,amsthm}
\usepackage{times}
\usepackage{graphicx}
\usepackage{subcaption}
\usepackage[ruled, noend,linesnumbered]{algorithm2e}
\usepackage[conf={no link to proof}]{proof-at-the-end}
\graphicspath{ {./figures/} }

\newtheorem*{thm*}{Theorem}

\newtheorem*{lemma*}{Lemma}

\newtheorem*{corollary*}{Corollary}

\newcommand{\clients}{n}

\title{Bit-efficient Numerical Aggregation and Stronger Privacy \\ 
for Trust in Federated Analytics}
\author{Graham Cormode, Igor L. Markov
\\ gcormode@fb.com, imarkov@fb.com
}

\begin{document}

\newcommand{\Exp}{\mathbb{E}}
\newcommand{\Var}{\mathbb{V}}

\newcommand{\hush}[1]{}
\newcommand{\red}[1]{{\color{red}#1}}
\newcommand{\blue}[1]{{#1}}
\newcommand{\eat}[1]{}
\maketitle

\begin{abstract}
 Private data generated by edge devices --- from smart phones to automotive electronics --- are highly informative when aggregated but can be damaging when mishandled.
 \eat{Secure aggregation, differential privacy, and other valuable solutions to this challenge}
 A variety of solutions are being explored but
 have not yet won the public's trust and full backing of mobile platforms. In this work, we propose numerical aggregation protocols that empirically improve upon prior art, while providing comparable local differential privacy guarantees. Sharing a single private bit per value supports privacy metering that enable privacy controls and guarantees that are not covered by differential privacy. We put emphasis on the ease of implementation, compatibility with existing methods, and compelling empirical performance.
 
 \eat{Our proposal is backed by aggregation protocols that are compatible with prior solutions, relatively straightforward to implement in practice,
 and amenable to formal analysis.}
\end{abstract}

\section{Introduction}

Smart phones and smart watches, fitness trackers, automotive electronics, building sensors, and other edge devices collect large amounts of private data, including locations, timestamps, behaviors, personal preferences, as well as data related to financial and medical information. While useful to online advertising, such data can be stolen and used maliciously, which has prompted serious concerns and mistrust in technology among the public, government regulators, and some industry participants.
A number of recent laws and regulations (e.g., {\tt https://gdpr-info.eu/}) limit collection or storage of private data, but a gap remains between what is allowed and what would be considered entirely acceptable by all parties involved. 
In particular, Apple has implemented more restrictive data collection policies in iOS 14.5 \cite{AppleIOS14}, whose impact on ad networks has been described in detail \cite{FacebookIOS14ads}. 
Learning to work with less private data than before is a challenge for existing ad networks. 
Convincing the public that their private data are safe is an even greater challenge.

 Technical solutions for improving data privacy are receiving significant attention from academia and industry, including Google (Chrome~\cite{Erlingsson2014}),
 Microsoft (Windows~\cite{Ding2017}), Apple (iOS~\cite{AppleDP2017}), Intel (SGX~\cite{IntelSGX}) and others. To limit the impact and potential damage caused by a private datum, aggregation is used whenever possible and often suffices. Aggregation can take the form of a sample mean, the 90th percentile or a histogram over a sufficiently large sample. 
 To protect data from leakage before aggregation is complete, Intel offers hardware with Secure Guard Extensions (SGX) \cite{IntelSGX}, which assumes trust in the security of hardware beyond an edge device and in communications with such hardware. 
 Another potential danger is that the aggregated data may nevertheless reveal some information about individual contributors in rare cases. 
 To prevent that, {\em federated analytics} methods \cite{GoogleFA2020} use aggregation protocols with mathematical guarantees (termed {\em differential privacy} \cite{Dwork2014} and extended to {\em federated learning} \cite{GoogleDP2017,AppleDP2017,AppleFed2021}), where the underlying idea is to add noise to private data before aggregation to provide {\em plausible deniability} through {\em randomized response} techniques~\cite{RandResponse1965}. 
 
Under differential privacy, any contribution of any client is compatible with any private values for that client, but some contributions are more likely. 
The strength of differential privacy is captured by the parameter $\epsilon > 0$ which (via $\exp(\epsilon)$) bounds the likelihood of
truthfully reporting about the correct data --- the lower the $\epsilon$, the stricter the privacy requirement.
{\em Local differential privacy} extends differential privacy guarantees to each client locally regardless of third-party servers \cite{LocalDP2011,LocalDP2020}, so that aggregators do not see original data. Sequential composition
\cite[Theorem 1]{LocalDP2020} enables more sophisticated algorithms.


Apple pioneered the use of differential privacy in its products \cite{AppleDP2017}, and the 2020 US Census~\cite{Haney2020} has been using differential privacy to aggregate results. These developments demonstrated that differential privacy can be deployed in serious applications but also exposed its technical shortcomings \cite{AppleDPpoor2017,USCensus2020} ---
counter-intuitive choices of privacy parameters, misunderstandings of differential privacy concepts, and deficient implementations impacted roll-out in practice. While promising and mathematically rigorous, differential privacy techniques {\em alone} do not currently appear compelling to the media \cite{AppleDPpoor2017}, the public, civic groups \cite{USCensus2020}, government regulators, the industry \cite{Levy2020} and even the research community~\cite{DPLimits}\cite[Section 7]{LocalDP2020}. 
\hush{
\begin{itemize}
\item Understanding differential privacy and specific techniques requires deep and diverse technical expertise,
\item The choice of the $\epsilon$ parameter for differential privacy is not intuitive and has been missing in public disclosures,
\item Reasonable tradeoffs between utility and privacy in terms of $\epsilon$ depend on domain-specific data distribution and on having a sufficiently large client base; they may be impossible in important cases.
\item Data correlations are not always accounted for in mathematical modeling.
\item The additional noise parameter $\sigma$ doubles the problems discussed above.
\item Fundamental misunderstandings of differential privacy and also its deficient implementations have jeopardized the roll-out of several industry and government solutions.
\end{itemize}
}
To put it plainly, differential privacy alone is insufficient in practice because it does not provide common-sense privacy guarantees that consumers and product managers tend to assume --- that an individual's data are safe from unauthorized parties (which are not included in the DP formalism).
Instead, an individual's data is shared for aggregation, thus shifting trust to the aggregation protocols, software and hardware used to implement aggregation. With its significant complexity, this strategy makes private data vulnerable to ($i$) future implementation mistakes ranging from hardware and software to security infrastructure and UIs, ($ii$) system glitches, and ($iii$) clever hacking along a variety of attack vectors.

\subsection{Our approach and contributions}

The mathematical sophistication of existing methods complicates audit and verification, increases the chances of implementation mistakes, and leaves many possible attack vectors. 
Our work addresses these challenges, while absorbing or extending prior solutions. In particular, we do not assume secure hardware but can make use of it. Likewise, we do not introduce novel differentially private mechanisms, but provide a DP guarantee by leveraging an off-the-shelf mechanism. 

\smallskip
\paragraph{Privacy metering.}
We propose to meter private data not at the value level (such as an integer representing someone's current longitude), but at the bit level. 
Rather than transmit an entire private value with noise added, our aggregation protocols only transmit a single private bit (and limit subsequent bits per value and per client).
From the perspective of a model like differential privacy, transmitting one bit out of 32 can be equivalent to adding noise on 31 bits and transmitting the entire value. 
However, actually transmitting fewer {\em private} bits simplifies audit, verification and metering. To this end, we show how to perform many types of aggregation by combining random bits from different edge devices. At a time when many federated analytics stacks are under active development, our proposal suggests that limits on transmitting private information can be surfaced as controls at the level of mobile platforms and, perhaps, help the public improve trust in technology.

\smallskip
\paragraph{Bit-efficient numerical aggregation.}
Our mathematical contributions are the introduction and analysis of more efficient techniques for the computational estimation of means, variances, etc. Prior LDP techniques either operate directly on real numbers, e.g., adding Laplace noise, or produce some noisy discrete values as a result of rounding, range checks, comparisons to thresholds, etc. In contrast, 
we approximate real numbers with fixed-point (or integer) representations, expand them in binary, select some of the resulting bits, and postprocess those bits before communicating them. Several considerations are both novel and essential to the correctness and efficiency of this approach:

\begin{itemize}
\item The bits we extract from the numbers to be aggregated form a {\em linear decomposition} of such numbers in each bit;\footnote{Note that {\em signed} binary expansions are not linear in the sign bit.}
\item Communicated bits have distinct weights associated with them (in the linear decomposition);
\item The sets of numbers that share a specified value of some bit do not usually form ranges, but do overlap.
\hush{In contrast to the histogram approach that uses equal-significance non-overlapping range-based buckets}
\end{itemize}
Additional considerations are helpful to simplify analysis, streamline implementation or improve performance:
\begin{itemize}
\item We sample bits using a custom-optimized frequency distribution;
\item Sampling frequencies can be adjusted based on early sampling results; in particular, unused bits (with estimated mean 0) do not need to be sampled;
\item Where genuine random sampling is unnecessary,
we replace it with quasi-Monte Carlo (QMC) methods.
\end{itemize}

 We show empirically that gathering statistics using the new binary-expansion approach offers significant advantages over established approaches as well
 as recently proposed techniques. 
Our analysis shows that the error decreases quickly, proportionally to $1/\sqrt{\clients}$, where $\clients$ is the number of client bit reports (formalized in Lemma~\ref{lem:basic}). 
In practice, we see that gathering reports from a few thousand users is sufficient to achieve a normalized RMSE of around 3\% for a 10-bit quantity, and ten thousand reports ensure that the error level is comfortably below 1\%.  
We observe that the efficiency of existing approaches often relies on knowing tight bounds on the range in which the values fall.  We relax this assumption and allow our approach to adapt to the distribution of values observed in practice. Additional optimizations and extensions are discussed below.
 
\subsection{Paper organization}
In the remaining part of the paper,
Section \ref{sec:prior} covers prior
work and identifies the main competitors
of our methods. Section \ref{sec:bitpushing} introduces our ``bit pushing'' technique for mean estimation, gives formal analysis and extends it to provide LDP guarantees. In addition to additive aggregation for which bit pushing was developed, we discuss nonlinear extensions, including variance estimation.
Section \ref{sec:experiments} introduces key implementation parameters, reports on parameter optimization, and compares the performance of bit pushing to that of leading prior techniques. 
Section \ref{sec:conclusions} puts our work in perspective, pointing out some limitations and pitfalls, discussing likely applications, and suggesting how follow-up efforts might proceed. In particular, novel privacy infrastructure should ensure the separation of private and non-private bits, and support privacy metering.

\section{Prior work}
\label{sec:prior}

When aggregating private values held by clients, a key challenge is to ensure sufficient privacy for plausible deniability. In particular, ($i$) the information shared by any client should be 
insufficient to deduce their private value, but ($ii$) the information shared collectively by many clients should suffice to estimate aggregate values with acceptable accuracy. Prior work approaches this challenge in terms of (Local) Differential Privacy~\cite{LocalDP2020}. 
Centralized Differential Privacy can add a relatively small amount of noise to the correct answer, but typically requires that each client disclose their true value. In this work, we prefer the stronger privacy constraint where each client reveals a smaller fraction of their data, via the local model of privacy. 

\smallskip

\paragraph{$\epsilon$-Local Differential Privacy}($\epsilon$-LDP or just LDP) requires that for any given output value that the client may emit, the probability of producing the same output for a different input can vary by at most a factor of $\exp(\epsilon)$. 
As $\epsilon$ approaches zero, this requires all outputs to approach uniform likelihood, whereas when $\epsilon$ tends to infinity, this imposes effectively no constraint on the output distribution. 
The baseline approach to differential privacy is to add appropriately scaled statistical noise to the true value of a function -- most commonly, noise sampled from the Laplace distribution (with variance $O(1/\epsilon^2)$).  
In our setting, this would entail each client taking their input value, and adding Laplace noise, scaled by the maximum value that a client may hold. 
This ensures that each report is masked effectively, but by aggregating reports from multiple clients, the independent noise will tend to cancel out on average, and impose an overall error that is an order of magnitude smaller than the aggregated signal. 

Adding Laplace noise is a simple, but rather blunt instrument. 
Several LDP mechanisms instead aim to add noise at the bit level, by making use of Randomized Response or its generalizations. 
In randomized response, a binary value is reported accurately with probability $p \ge \frac12$, else its complement is sent. 
This achieves LDP with $\epsilon = \ln(p/(1-p))$. 
A simple way to incorporate randomized response is to combine it with randomized rounding, as proposed by Duchi \textit{et al.} in an early work on LDP~\cite{Duchi2018}.  
Assuming that an input value $x$ is pre-scaled so that $0 \le x \le 1$, we can treat $x$ as a probability, and represent it with 1 with probability $x$, else with 0 ({\em randomized rounding}).
Applying randomized response to this bit satisfies the LDP requirement. Gathering many such noisy binary reports from clients, we can compute an unbiased estimate of the population mean. The variance of each response is $O(\frac{(\exp(\epsilon)+1)^2}{(\exp(\epsilon)-1)^2})$.
Similar ideas have been deployed by Microsoft for Windows app usage data collection~\cite{Ding2017}. 
Subsequent work by Wang \textit{et al.} modifies the procedure to sample a value so that values closer to the input
are chosen with higher probability than those that are further away, referred to as the ``piecewise'' mechanism~\cite{Wang2019}.  
Another common approach to privacy represents a value via a histogram: given a set of bucket divisions of the input space, we effectively represent an input value as a one-hot vector.  
For privacy, each client noises their vector, by perturbing it according to an LDP mechanism, and shares the noisy result with an aggregator~\cite{Wang2017}. 
The error in the estimate comes from combining the noise added (which grows proportionally to the number of buckets) and the uncertainty due to placing values in buckets (which decreases with the number of buckets). 
Balancing these two effects when estimating the mean of $\clients$ reports leads to variance proportional to $O(\frac{1}{\epsilon \sqrt{\clients}})$. 
%

\smallskip

\paragraph{Communication-efficient numerical aggregation} (regardless of privacy)
can be illustrated by theoretical work on low-bandwidth sensor networks~\cite{Luo2005} with 
``the constraint that the communication from each sensor to the fusion center must be a one-bit message.'' Here, a natural approach to gather 
statistics on distributed data
is to sample individual bits.
However, when dealing with packet-switched networks such as TCP/IP (as in our work), the basic transmission unit is a {\em packet} of at least 100-1000 bytes, and optimizing communication below that level is not a direct concern. Despite different objectives and resource metrics, basic analysis by Luo~\cite{Luo2005} concords with ours: in the absence of further information about the data distribution, sending binary-expansion digits with exponentially decreasing probabilities is the optimal strategy. From there, our studies diverge since Luo analyzes undesirable uncontrolled noise on sensor readings, whereas we deliberately add carefully calibrated bit-level noise to values that are being aggregated, in order to attain local differential privacy guarantees.

More recent interest in communication-efficient transmission of values (also not tied to privacy) has been motivated by machine-learning applications. Dealing with multi-dimensional data, distributed ML applications may be able to leverage bit-level efficiency to reach-packet efficiency.
 Ben-Basat \textit{et al.} analyze leading approaches for estimating a real value using a single bit sent from a client to a server, as a function of the amount of shared randomness between the two parties~\cite{Ben-Basat2020}. 
 In our setting, the relevant point of comparison is given by {\em subtractive dithering}.\footnote{When we evaluated in our setting several approaches that were described in~\cite{Ben-Basat2020}, subtractive dithering was a clear frontrunner.}
For their input value $0 \leq x \leq 1$, each client samples $h \in U[0,1]$, and send $b = 1$ if $x\ge h$. 
The server receives $b$ and $h$, and estimates $\hat{x} = b + h - 0.5$. 
This offers some privacy when $h\approx\frac12$, but when $h$ is chosen to be close to 0 or 1, we can learn if the value of $x$ is close to $h$. 
The variance for each estimate (between 0 and 1) is bounded by a constant. In order to use subtractive dithering as a baseline for comparisons to our work, we add a formal privacy guarantee by applying randomized response to the input-dependent output $b$. 

\smallskip
\paragraph{The need for adaptive protocols.}
The methods above assume inputs in the range $[0, 1]$ or,
equivalently, in some range $[L, H]$ so they can be mapped to $[0, 1]$ via $f(x) = \frac{x - L}{H - L}$. 
Assuming loose bounds on the input values has a negative impact on accuracy: while methods that are optimal for $[0,1]$ can be applied to $[L, H]$, the variance of their estimates scales with $(H-L)^2$~\cite{Ben-Basat2020}. 
More promising are protocols that adapt to the data distribution and ``zoom in'' on the range where the data truly lies. 
We implement such adaptation in our protocols using a small number of rounds and show ($i$) sharper analytical bounds for variance, backed by ($ii$) reduced variance in simulations.
When combined with the intuitive nature of the bit-level privacy metering and ease of achieving a formal $\epsilon$-LDP guarantee, bit-pushing becomes an attractive option for mean estimation and related tasks. 
\blue{Our approach to localizing the range of the data could also be combined with other methods to estimate the mean within the resulting range. 
The advantage of the bit-pushing approach to find the data range is that it operates with only a single round of interaction, rather than the multiple rounds that would be required to perform a binary search. 
}

\section{The Bit-pushing Approach}
\label{sec:bitpushing}

In this section, we outline our algorithms for mean estimation, and formally analyze their properties. They can be used as subroutines in many applications including federated learning and can be extended to aggregate some nonlinear quantities.

\subsection{Basic bit pushing algorithm}
\label{sec:bitpushingbasic}
Assume that each client $i$ out of $\clients$ owns a private value $x_i$: we work with $b$-bit integer and fixed-point values.
In the narrative below, we first assume non-negative integers, but this is not a limitation of the approach. Our goal is to estimate the mean $\bar{x} = \sum_{i=1}^\clients (x_i/\clients)$. 
We write $x^{(j)}$ to denote the $j$'th bit in the binary representation of $x$, and $\bar{x}^{(j)}$ to denote the $j$'th bit of the mean, $\bar{x}$. 
In the basic form of bit pushing, each client selects bit $j$ with probability $p_j$, and sends the value of their input at this bit location, as the pair $\langle x_i^{(j_i)}, j_i \rangle$. 
An alternative is where the server randomly selects a $p_j$ fraction of clients to report back on bit $j$.  
This reduces the variance in the number of reports of each bit, and removes the requirement for the server to specify a sampling distribution to clients.  
Unless stated otherwise, we adopt this quasi-Monte Carlo sampling method as the default. 

\begin{theoremEnd}{lemma}
The basic bit pushing protocol provides an estimate that is unbiased and has variance equal to
$\frac{1}{\clients} \sum_{j=0}^{b-1} \frac{4^{j} \bar{x}^{(j)}(1 - \bar{x}^{(j)})}{p_j}$.
\label{lem:basic}
\end{theoremEnd}
\begin{proofEnd}
Let $X^{(j)}$ denote the distribution of the $j$'th bit value.  
We can assume that each $X^{(j)}$ follows a Bernoulli distribution with parameter $\Exp[X^{(j)}]$. 
Assuming the quasi-Monte Carlo case, where
bit $j$ is reported on by exactly $\clients p_j$ clients, our estimate 
$\hat{X}^{(j)}$ is the mean of these $\clients p_j$ reports. 
Clearly $\Exp[\hat{X}^{(j)}] = \Exp[X^{(j)}]$, which, by linearity of expectation, is $\bar{x}^{(j)}$. 
Our estimate $X$ is the sum of these bit means, weighted by $2^j$, 
so $X = \sum_{j=0}^{b-1} 2^j \hat{X}^{(j)}$ (applying linear decomposition), 
and by definition, 
\begin{equation}\Exp[X] = \Exp\left[\sum_{j=0}^{b-1} 2^j X^{(j)}\right] = 
\sum_{j=0}^{b-1} 2^j \bar{x}^{(j)} = \bar{x} . 
\end{equation}

For bit $j$, each report on this bit is assigned weight $2^j$. 
The corresponding contribution to the variance is 
$\Var[2^j X^{(j)}] = 4^j \bar{x}^{(j)}(1 - \bar{x}^{(j)})$. 
Averaged over the $\clients p_j$ reports, the contribution to the variance from the estimate of bit 
$j$ is $4^j \bar{x}^{(j)}(1 - \bar{x}^{(j)})/(\clients p_j)$, so the overall variance of the estimator
is 
\begin{equation}
    \Var[X] = \sum_{j=0}^{b-1} \frac{4^j}{\clients p_j} \bar{x}^{(j)}(1 - \bar{x}^{(j)}) := \frac{1}{\clients} \sum_{j=0}^{b-1} \frac{\beta_j}{p_j}
    \label{eq:var}
\end{equation}
\end{proofEnd}
%
%
%
\begin{theoremEnd}{corollary}
If each client sends $b_\mathrm{send}$ bits, the variance decreases to 
$\frac{1}{\clients b_\mathrm{send}} \sum_{j=0}^{b-1} \frac{4^{j} \bar{x}^{(j)}(1 - \bar{x}^{(j)})}{p_j}$
\end{theoremEnd}

\begin{proofEnd}
This result follows immediately by adaptive the previous proof to average over 
$(\clients p_j b_\mathrm{send})$ samples for bit $j$.  
\end{proofEnd}

The proofs of all claims are presented in the Appendix. 
This result relies on the fact that the mean is a linear function of the values at each bit location. 
Clearly, the quality of this bound will depend on the choice of the sampling probabilities $p_j$, so we consider different choices for setting the $p_j$ values. 
We note that it is likely that individual bits will be correlated -- for example, if the mean is close to a power of 2, then it is much more likely that the top-two most significant bits of the input will be $01$ or $10$ than $11$ or $00$. 
This does not affect our variance bounds. 
If $b_{\mathrm{send}} = 1$, then such correlations do not impact the protocol, since each sampled input value is independent of the others. 
Meanwhile, if $b_{\mathrm{send}} > 1$, then bit correlations will have negative covariance, and so reduce the variance of our estimates further. 
In what follows, we focus on the case $b_{\mathrm{send}} = 1$. 
Algorithm~\ref{alg:bp} gives pseucocode for the core functionality of bit pushing, given a probability vector $p$ of weights to sample bits with. 

\smallskip
\noindent
{\bf Uniform sampling probabilities.}
The simplest setting is to pick $p_j = 1/b_{\max}$, i.e., each bit is uniformly likely to be picked. 
In this case, our bound on the variance becomes $\frac{b_{\max}}{\clients}  \sum_{j=0}^{b_{\max}-1} \bar{x}^{(j)}(1-\bar{x}^{(j)}) 4^j \le \frac{b}{3\clients} 4^{b_{\max}}$ (since
$x(1-x) \leq \frac14$ for $0 \leq x \leq 1)$.
Commonly, we may expect that the mean value is proportional to $2^{b_{\max}}$ (i.e., when $\bar{x}^{(b_{\max})}$ is a constant). 
Then we can write the variance as proportional to 
$b_{\max} \bar{x}^2/\clients$, and the expected absolute error to be proportional to $\sqrt{b_{\max}} \cdot \bar{x}/\sqrt{\clients}$. 
However, this is a suboptimal choice, and we can do strictly better as discussed below. 
Increasing $b_\mathrm{send}$ from 1 towards $b_{\max}$ would have the effect of removing the dependence on $b_{\max}$ in the variance, although 
when $b_\mathrm{send} = b_{\max}$, the algorithm is equivalent to sending the client's entire input value. 

\smallskip
\noindent
{\bf Weighted sampling probabilities.}
It seems intuitive that higher-order bits should have greater probability of being sampled, since they contributed more highly to the computation. 
Some natural choices are $p_j \propto 2^{j}$, or more generally, $p_j \propto c^j = 2^{\alpha j}$ for some $c$ or $\alpha$. 
We see next that this is indeed a principled choice. 

\smallskip
\noindent
{\bf Optimizing sampling probabilities.}
To look for a trend, we can treat the input bits as independent.

\begin{theoremEnd}{lemma}
  The variance of the bit-pushing estimator is minimized by picking
  $p_j = \sum_{j=0}^{b-1} \frac{\sqrt{\beta_j}}{A}$, 
  where $\beta_j = 4^{j} \bar{x}^{(j)}(1 - \bar{x}^{(j)})$, 
  and $A = \sum_{j=0}^{b-1} \sqrt{\beta_j}$. 
\label{lem:minvar}
\end{theoremEnd}

\begin{proofEnd}
For a fixed budget of bit samples, we seek to minimize
$\Var[X] = \frac{1}{\clients} \sum_j \frac{\beta_j}{p_j}$ with $\beta_j, p_j > 0$. 
To optimize variance in terms of $p_j$ such that $\sum_j p_j = 1$, we perform unconstrained optimization of
\begin{equation}
f(p_1,\ldots,p_{k-1}) =
\frac{\alpha_k}{1 - \sum_{j=0}^{k-1}p_j} + \sum_{j=0}^{k-1} \frac{\beta_j}{p_j}
\end{equation}

Looking for a local extremum inside the probability simplex, we obtain

\begin{equation}
\forall i = 1 .. k - 1, ~~~
\frac{\partial f(p_1,\ldots,p_k)}{\partial p_i} = \frac{\alpha_k}{(1 - \sum_{j=0}^{k-1}p_j)^2} -\frac{\alpha_i}{p_i^2} = 0
\end{equation}

Therefore
\begin{equation}
\forall i,l \quad
\frac{\alpha_i}{p_i^2} = \frac{\alpha_l}{p_l^2}
\qquad \Rightarrow \qquad
 p_i / p_l = \sqrt{\alpha_i/\alpha_l}
\end{equation}
and this extends to $l=k$ via renumbering. Therefore,
$p_j = \sqrt{\beta_j} \frac{p_k}{\sqrt{\alpha_k}}$, and to find $p_j$, we can just $L_1$-normalize the vector of $\sqrt{\beta_j}$. To confirm that this unique critical point is the global minimum, we compute

\begin{equation}
\forall i, j ~~~
\frac{\partial \Var}{\partial p_i \partial p_j} =
\left\{
 \begin{array}{lr}
  0, & \mathrm{for}~ i \neq j \\
  2 \beta_j / p^3_j \clients & \mathrm{for}~ i = j \\
 \end{array}
\right\}
\end{equation}
Given that $p_j, \beta_j > 0~\forall j$, the Hessian is positive semidefinite. 
\end{proofEnd}

From Lemma~\ref{lem:basic} and our independence assumption, we have 
$\beta_j = \bar{x}^{(j)}(1-\bar{x}^{(j)}) 4^j$, where $\bar{x}^{(j)}$ denotes the mean value at the $j$'th bit index. 
If we simply bound the contribution of $\bar{x}^{(j)}(1-\bar{x}^{(j)})$ values by $\frac14$, so that $\beta_j = \frac14 4^j$, this leads us to set $p_j = 2^{j}/(2^b -1)$, and so we obtain
\begin{equation}
    \Var[X] \leq \frac{1}{\clients}  \sum_j (2^{b}-1) \bar{x}^{(j)}(1-\bar{x}^{(j)}) 2^j < \bar{x} (2^{b-2})/\clients .
    \label{eq:bp-var}
\end{equation}

If the inputs make use of most input bits, then $\bar{x}$ is reasonably large, i.e., $\bar{x} \propto 2^b $, and so 
$\Var[X] \propto \bar{x}^2/\clients$. 
Hence, the expected absolute error will be of magnitude $\bar{x}/\sqrt{\clients}$.
Sending $b_{\mathrm{send}}> 1$ bits per client would further reduce this absolute error by a factor of 
$1/\sqrt{b_{\mathrm{send}}}$. 

\paragraph{Local vs. central randomness.}
On the surface, it does not matter whether the choice of which bit(s) to sample is performed by each client (local randomness) or prescribed by the server to each client (central randomness). 
In other words, rather than sending bit-sampling frequencies to each client, the server can send the bit index/indices to sample, potentially reducing communication costs. 
Given that bit-sampling frequencies tend to be skewed toward significant bits, letting the client to sample bits does not significantly improve privacy.
However, the local setting is more vulnerable to clients who may try to poison the response by distorting the reported values of high-order bits.
On the other hand,
central randomness supports quasi-Monte Carlo sampling, where the server enforces sampling
frequencies, potentially improving the accuracy of estimation. 
Therefore, we favor central randomness.

\begin{algorithm}[t]
\DontPrintSemicolon
\SetKwInOut{Input}{Input}\SetKwInOut{Output}{Output}
\Input{No. of bits $b$, bit weights $p$, no. of clients $\clients$} 
\Output{ (Result $r$, mean of bits $m$, sum of bits $s$)}
Initialize result $r=0$ \;
\For {$j = 0$ \textup{\textbf{to}} $b-1$ \textup{\textbf{in parallel}}}{
   Contact $c[j] = p[j]\cdot \clients$ clients to request bit $j$ \;
   Gather weighted sum of bits as $s[j]$ \;
   Compute bit means $m[j] = s[j]/c[j]$ \;
   $r \gets r + (2^j \cdot m[j])$ \;
}
\Return $(r, m, s)$
\caption{Basic bit-pushing algorithm\label{alg:bp} }
\end{algorithm}

\begin{algorithm}[t]
\DontPrintSemicolon
\SetKwInOut{Input}{Input}\SetKwInOut{Output}{Output}
\Input{No. of bits $b$, no. of clients $\clients$, parameters $\alpha$, $\gamma$, $\delta$} 
\Output{ Result $r$}

\tcc{Round 1:}
\For {$j=0$ \textup{\textbf{to}} $b-1$}{
  Compute  $p_1[j] = (2^j)^\gamma$}
Normalize $p_1$: $p_1 \gets p_1 / \text{sum}(p_1)$ \; 
Run basic bit pushing: $(r_1, m_1, s_1) = \texttt{BitPushing}(b, p_1, \delta \clients)$\;
\tcc{Round 2:} 
\For {$j = 0$ \textup{\textbf{to}} $b-1$}{
  Compute $p_2[j] = (4^j \cdot m_1[j] \cdot (1-m_1[j])^\alpha$
 }
Normalize $p_2$: $p_2 \gets p_2 / \text{sum}(p_2)$ \; 
Run basic bit pushing: $(r_2, m_2, s_2) = \texttt{BitPushing}(b, p_2, (1-\delta)\clients)$ \;
\tcc{Final aggregation:}
Combine means $m_3 = (s_1 + s_2)/(\delta \clients * p_1 + (1-\delta)\clients * p_2)$ \;
\For {$j = 0$ \textup{\textbf{to}} $b-1$}{
  $r \gets r + 2^j \cdot m_3[j]$
}
\Return $r$
\caption{Adaptive bit pushing algorithm\label{alg:abp} }
\end{algorithm}

\subsection{Adaptive Bit Pushing}
\label{sec:adaptivebitpushing}

A more sophisticated approach is to use a first round of bit pushing to estimate the bit means $\bar{x}^{(j)}$. 
That is, we first choose a set of sampling probabilities $p_j$ independent of the input, and ask a $\delta$ fraction of the clients to report an input bit according to this distribution. 
From these reports, we estimate $\bar{x}^{(j)}$ as $\hat{x}^{(j)}$ for all $j$, and use these estimates to compute a new set of weights based on $\beta'_j = \hat{x}^{(j)} (1-\hat{x}^{(j)})4^j$. 
We can then perform a second round of bit pushing using sampling probabilities $p_j = \sqrt{\beta'_j}/\sum_{j=0}^{b-1} \sqrt{\beta'_j}$ for the remaining $1-\delta$ fraction of clients. 
To instantiate this two-round approach, we need to determine
($i$) what split parameter $\delta$ to apply; and
($ii$) how to choose the initial weights $\beta_j$. 
Naively, we might choose $\delta = \frac12$ to balance accuracy of learned $\beta'_j$s and accuracy of reported results.
For $\beta_j$s, we might default to choosing $\beta_j = 4^j$ (and hence $p_j \propto 2^j$), according to the above argument. 
Our analysis below guides the choice of $\delta$, and we will try different settings for both these choices in our empirical evaluations. 
Algorithm~\ref{alg:abp} provides pseudocode for the adaptive bit pushing approach, where $*$ is used to denote scalar-vector multiplication. 

\paragraph{Analysis of adaptive bit pushing.}
We seek to analyze the variance of two-round adaptive bit pushing.

\begin{theoremEnd}{lemma}
The variance of adaptive bit pushing is bounded by $\frac{b_{\max} \sigma^2}{\clients} + O(b_{\max} 4^{b_{\max}}/\clients^{3/2})$, where
$b_{\max}$ is the index of the highest-order bit that is non-zero in the input.
\end{theoremEnd}

\begin{proofEnd}
With adaptive bit pushing, the first round allows us to find accurate estimates for each of the $\beta_j$ parameters in \eqref{eq:var}, and
we proceed by substituting our choice of $p_j$ into \eqref{eq:var}.
We first assume that the estimates from the first round give exact values for $p_j \propto \sqrt{\beta_j}$. 
Write $A = \sum_{j=0}^{b-1} \sqrt{\beta_j}$, so that $p_j = \sqrt{\beta_j}/A$. 
Then \eqref{eq:var} sets
\begin{equation}
    \Var[X] = \frac{1}{(1-\delta)\clients} \sum_{j=0}^{b-1} A \frac{\beta_j}{\sqrt{\beta_j}} = \frac{A}{(1-\delta)\clients} \sum_{j=0}^{b-1} \sqrt{\beta_j} = 
\frac{A^2}{(1-\delta)\clients}
\label{eq:bp-abp}
\end{equation}

Next, we can observe that, using the Cauchy-Schwarz inequality on a vector of up to $b_{\max}$ different $\beta_j$ values,  
\begin{equation}
A^2 \leq b_{\max} \sum_{j=0}^{b_{\max}-1} \beta_j = b_{\max} \sum_{j=0}^{b_{\max}-1} 4^j\bar{x}^{(j)}(1 - \bar{x}(j)) := b_{\max} \sigma^2
\label{eq:abp-A}
\end{equation}

\noindent
where, $\sigma^2$ is the variance of the input distribution since, by linearity of expectation, 
we can decompose $\sigma^2 = \sum_{j=0}^{b-1} (\sigma^2)^{(j)} = \sum_{j=0}^{b-1} 4^j\bar{x}^{(j)}(1 - \bar{x}^{(j)})$. 
Hence, our estimate is as efficient as the trivial (non-private) estimator of taking the mean of $n$ samples, up to the factor of 
$b_{\max}/(1-\delta)$. 

We now consider the effect of sampling $\delta \clients$ clients in the first round. 
For a given bit $j$, if we estimate this based on a sample of size $s$, then the error in our estimate of $\bar{x}^{(j)}$ will be proportional to 
$1/\sqrt{s}$, from standard sampling bounds. 
Then our estimate of $\beta_j$ (as $\hat{\alpha}_j$) will accordingly have an error of $O(4^j/\sqrt{s})$. 
Putting this into our expression for $\sigma^2$ yields 
\begin{equation}
    \hat{\sigma}^2 = \sum_{j=0}^{b_{\max}} \hat{\alpha}_j = \sum_{j=0}^{b_{\max}} (\beta_j + O(4^j/\sqrt{s})) < \sigma^2 + O(4^{b_{\max}} \frac{1}{\sqrt{s}}).
    \label{eq:abp-error}
\end{equation}
Hence, the contribution to the total variance from this error term is, from~\eqref{eq:bp-abp}, \eqref{eq:abp-A}, and \eqref{eq:abp-error}, 
proportional to $\frac{ 4^{b_{\max}}}{(1 - \delta)\clients \sqrt{\delta \clients}}$. 
This term is minimized, as a function of $\delta$, if we maximize the expression
$(1 - \delta)\sqrt{\delta}$. 
Writing $z = \sqrt{\delta}$, we aim to maximize
$z - z^3$. 
We differentiate to obtain $1 - 3z^2 = 0$, and so set $z^2 = \delta = 1/3$.

Consequently, our bound on the variance is
$b_{\max}\sigma^2/\clients + O(b_{\max} 4^{b_{\max}}/\clients^{3/2})$. 
Provided the variance in the high-order bits is significant (e.g., if $(\sigma^2)^{(b_{\max})}$ is at least a constant), 
then this is dominated by the first term of $\sigma^2/\clients$. 
\end{proofEnd}

\paragraph{Comparison to alternate approaches.}
The benefit of adaptive bit pushing can be most easily understood when we have only a loose estimate of $b$, the number of bits to represent the input values. 
For methods which scale the input down to the range $[0,1]$ and then scale the estimated fraction back up, the variance of the resulting estimate is proportional to $(2^b)^2/\clients$. 
For (non-adaptive) bit pushing, it is proportional to $2^b\bar{x}/\clients$, as shown in \eqref{eq:bp-var}. 
Since adaptive bit-pushing allows us to identify any bits $j$ with $\bar{x}_j = 0$,  
we can bound the variance of the estimate by $2^{b_{\max}}\bar{x}/\clients$,\footnote{Note that 
the dependence on $2^{b_{\max}}$ is a somewhat unintuitive quantity, and arises from pessimistically assuming that sampling weights are proportional to $2^{j}/(2^{b_{\max}})$.} 
or use the above analysis to argue that the variance is proportional to
$b_{\max} \sigma^2/\clients$ plus lower order terms. Compared to the bound  \eqref{eq:bp-var} for our non-adaptive protocol, variance is reduced by a factor of $2^{b - b_{{\max}}}$.

\paragraph{Caching.}
We note a simple but impactful optimization.
Rather than producing our final estimate based solely on the reported values from the second round, we can pool the reports from both rounds. 
The net effect will be to gain more reports for each bit index, which should only improve the observed accuracy. 
We will also evaluate this ``caching'' technique in our experiments. 

\subsection{Local Differential Privacy Variant}

Given a (private) bit $y$, \textit{randomized response} is a simple procedure to mask the bit: with (public) probability $p$, we report $y$, otherwise we report $\bar{y} = 1-y$. 
To unbias this report, we replace a reported value $r$ with $\frac{r - (1-p)}{2p-1}$.  
If we run this procedure with $p = \frac{\exp{\epsilon}}{1 + \exp{\epsilon}}$, the mechanism achieves the $\epsilon$ (local) differential privacy guarantee. 
It is straightforward to give bit pushing an $\epsilon$-LDP guarantee: we apply randomized response to each bit before it is sent, and unbias the results at the server side. 
The variance of this unbiased estimator is $\frac{\exp{\epsilon}}{(\exp{\epsilon} - 1)^2}$. 
In contrast to the above analysis, this variance is independent of the bit means $\bar{x}^{(j)}$. 
When we apply randomized response to bit pushing, where we assign $p_j n$ clients to report on bit $j$, which is scaled by $2^j$, then we obtain a total bound on the variance of $\sum_{j = 0}^{b-1} \frac{4^j}{p_j \clients} \frac{\exp{\epsilon}}{(\exp{\epsilon} - 1)^2}$.
This is optimized according to the above argument by choosing $p_j = 2^j/(2^b-1)$, which yields a total variance bound of $O(\frac{4^b}{\clients} \frac{\exp{\epsilon}}{(\exp{\epsilon} - 1)^2})$. 

Note that for small $\epsilon < 1$, this expression is $O(\frac{4^b}{\clients} \cdot \frac{1}{\epsilon^2})$, and so the expected absolute error is $O(2^b/\epsilon\sqrt{\clients})$.

\subsection{Nonlinear Estimation}

Bit pushing is a natural fit to additive aggregation, such as mean estimation, because the binary expansion of non-negative integers is a linear function of individual bits. 
Hence, the question whether non-linear aggregation also benefits from bit pushing.
In this section, we describe how the approach can be extended to several non-linear estimation tasks.

\paragraph{Signed values.}
Consider the additive aggregation of signed integers, whose value depends nonlinearly on the sign bit. 
We consider two ways to address this task. 
First, this case can be handled by adding a large number $C$ so as to make each client value positive during the aggregation, 
and then subtracting $C$ times the number of clients $\clients$ from the aggregate.
This is easy to implement, but 
requires a good bound on $C$, the shift value. 
This {\em additive shifting} approach can be contrasted to what we call {\em bit-splitting}. For each bit $b_i$, we consider two derived bits computed as {\tt AND(sign==0, $b_i$)} and {\tt AND(sign==1, $b_i$)}. 
This way, we separately aggregate positive and negative values and then add the estimates of their sums. 
In our initial simulations, there is no clear winner when a tight bound on $C$ is known, but when only a loose bound is available, the signed approach outperforms the shifting approach. 

\paragraph{Variance estimation.}
The empirical variance of a collection of data items is another fundamental aggregate. 
Among many other applications, having estimates of the mean and the variance immediately enables {\em feature normalization} in federated learning. 
Computing the variance of the inputs, $\sigma^2$,
can be reduced to two mean estimations of derived values: $\Var[X] = \Exp[(X - \Exp[X])^2] = \Exp[X^2] - (\Exp[X])^2$. 
These two expressions are equivalent when evaluated exactly but behave differently when used for estimation.

\begin{theoremEnd}{lemma}
The estimated variance using bit-pushing via $\Exp[(X - \Exp[X])^2]$
has variance proportional to $(\sigma^2 + \bar{x}^2/\clients)^2/\clients$, 
while the estimated variance using bit-pushing via $\Exp[X^2] - (\Exp[X])^2$
has variance proportional to $(\sigma^2 + \bar{x}^2)^2/\clients$. 
\end{theoremEnd}

\begin{proofEnd}
\textbf{Case 1: $\Exp[(X - \Exp[X])^2]$.}
Assume we have an (unbiased) estimator for $\Exp[X] = \bar{x}$, which gives us 
$\hat{{x}}$.  
Let the variance of this estimator be denoted by $\bar{\sigma}^2_x$. 

Let $Z = (X - \hat{{x}})^2$, and we seek to approximate the mean of $Z$. 
We first compute $\Exp[Z]$:
\begin{align*}
    \Exp[Z] & = \Exp[(X - \hat{{x}})^2] \\
    & = \Exp[(X - \bar{x} + \bar{x} - \hat{{x}})^2] \\
    & = \Exp[(X - \bar{x})^2 + (\bar{x} - \hat{{x}})^2 + 2(X-\bar{x})(\bar{x} - \hat{{x}}) ] \\
    & = \Exp[(X - \bar{x})^2] + \Exp[(\bar{x} - \hat{{x}})^2] + 2\Exp[(X - \bar{x})(\bar{x} - \hat{{x}}) ]\\
    & = \sigma^2 + \bar{\sigma}^2_x + 0
\end{align*}

From our analysis of bit pushing above (Section~\ref{sec:bitpushingbasic}), the variance of the estimate using bit pushing 
can be approximated by 
$\bar{z}^2/\clients$, 
i.e., $(\sigma^2 + \bar{\sigma}_x^2)^2/\clients$. 
Similarly, we can bound $\sigma_x^2 \propto \bar{x}^2/\clients$. 
So the total variance is approximated by $(\sigma^2 + \bar{x}^2/\clients)^2/\clients$.

\smallskip
\noindent
\textbf{Case 2: $\Exp[X^2] - (\Exp[X])^2$.}
As above, assume we have $\hat{{x}}$ with variance $\bar{\sigma}^2_x$ (approximated by $\bar{x}^2/\clients$). 
We will use our bit-pushing estimator (twice) to estimate $\Exp[X^2]$ and $(\Exp[X])^2$, then subtract the two to get the final estimate of $\Var[X]$.

If we let $Y = X^2$, then using the analysis above (Section~\ref{sec:bitpushing}), we can bound the expected squared error in our estimate of $Y$ (as $\hat{{y}}$)
by $\bar{y}^2/\clients$. 
We can use the fact that $\Var[X] = \Exp[X^2] - (\Exp[X])^2$ to write 
$\bar{y} = \Var[X] + \bar{x}^2$, and so 
the error in estimating $\Exp[X^2]$ is bounded by
$(\sigma^2 + \bar{x}^2)^2/\clients$. 
We will overlook the contribution to the squared error from approximating the second term ($(\Exp[X])^2$), although this may also be large. 


\smallskip
To compare standard deviations,
we get $(\sigma^2 + \bar{x}^2/\clients)/\sqrt{\clients}$ for Case 1, 
and $(\sigma^2 + \bar{x}^2)/\sqrt{\clients}$ for Case 2.
Consequently, the Case 1 estimator is more attractive. 
\end{proofEnd}


\paragraph{Higher moments.}

The same approach can be adopted to estimate higher moments. 
As in the variance case, care is needed to ensure numerical stability of the estimates, and more bits are needed: the
$k$th moment will in general require $kb$ bits to represent. 
Higher moments provide one way to track the overall shape of the distribution~\cite{Gan2018}. 


\paragraph{Estimating products and geometric means.}
In a simple example, we can compute products by applying additive aggregation to logarithms computed by clients and then exponentiating the aggregate at the server.  
Here, more care is needed to ensure that sufficient accuracy is maintained for the values in their logarithmic form, for example by tracking additional bits of precision below the decimal point. 
This allows finding geometric means via the mean of the logarithms. 

\eat{
\subsection{Alternative: randomized rounding}

A standard \red{alternative approach~\cite{}} is to apply randomized rounding to each input, and optionally applying randomized response. 

That is, given $b$-bit input (non-negative) integer $x$, we compute the probability $p_x = x/2^b$. 

Then we output 1 with probability $p_x$, and 0 otherwise.  
It is immediate that the expected value of this procedure is $p_x$, and hence we can recover an estimate of $x$ by rescaling by $2^b$. 

To achieve LDP, we just have to additionally apply randomized response to the output bit, and unbias the received reports. 

\subsection{Alternative: histogram encoding}

- Divide the range $0\ldots 2^b$ into uniform ranges. 

- Report which bucket each value falls into. 

- Optionally, add randomized response noise to each bucket.

Not very accurate. 
}

\section{Experiments}
\label{sec:experiments}
We perform exeriments based on implementations in Python, and simulate the computational costs and accuracy of mean estimation.  
In addition to the bit-pushing approach, we also implement methods based on 
Laplace noise addition (``Laplace''), 
piecewise-constant output distribution~\cite{Wang2019} (``piecewise''), 
randomized-rounding, 
and subtractive dithering~\cite{Ben-Basat2020} (``dithering''). 
We compare these to our two main variants of bit pushing: the two-round adaptive bit pushing (``adaptive''), and the single round approach based
on a fixed allocation of weights to bits (``weighted''). 

\subsection{Experimental Setup}

We perform experiments on both human-generated and synthetic data. To generate synthetic data, we draw values from Normal, uniform and exponential distributions with varying parameters, as specified below. 
The human-generated data ({\em real data} for short) are the reported ages from publicly-available deidentified data from the US Census~\footnote{\url{https://archive.ics.uci.edu/ml/datasets/Census-Income+\%28KDD\%29}}.


Our main focus is on comparing the accuracy of the different techniques, measured by the normalized root-mean-squared error (NRMSE): 
in each experiment, we compare the true (empirical) value of the mean $\mu$ to the estimate $\hat{x}$, and compute
the mean of the squared difference over 100 independent repetitions, then divide by the true mean $\mu$ for normalization. 
When error bars are shown on our plots, they indicate the standard error of these repetitions. Our default number of clients --- 10K --- is representative of typical federated analytics scenarios.

\begin{figure*}[t]
\centering
\subcaptionbox{Varying second-round parameter $\gamma$ \label{fig:bp_gamma}}{\includegraphics[width = 0.33\textwidth]{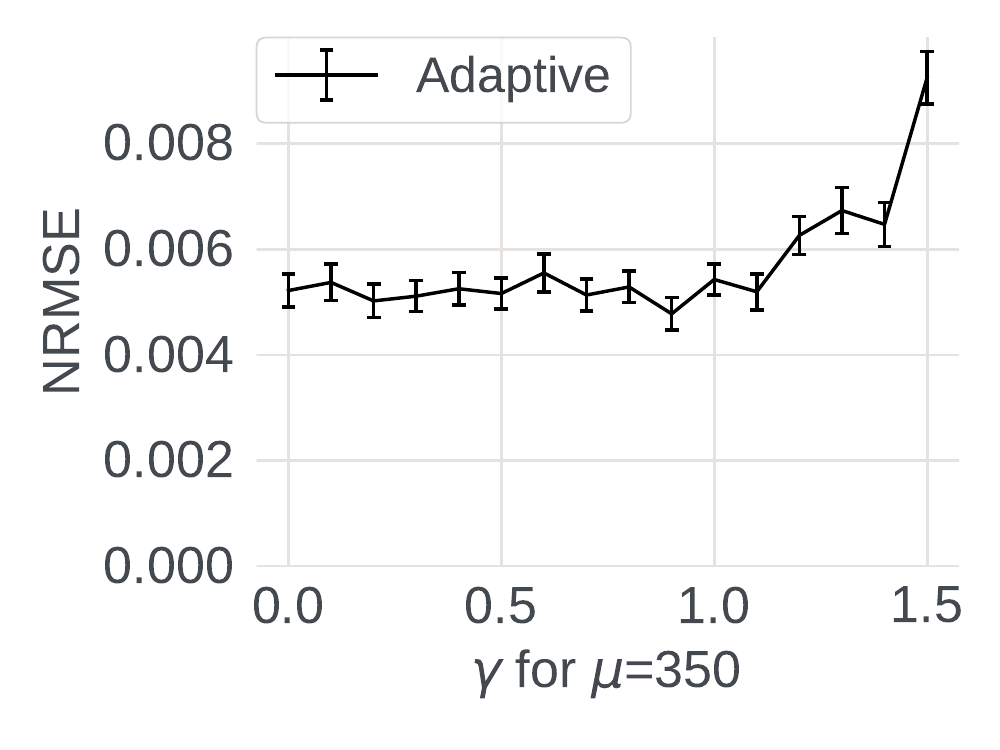}}%
\subcaptionbox{Varying adaptive parameter $\delta$ \label{fig:bp_delta}}{\includegraphics[width = 0.33\textwidth]{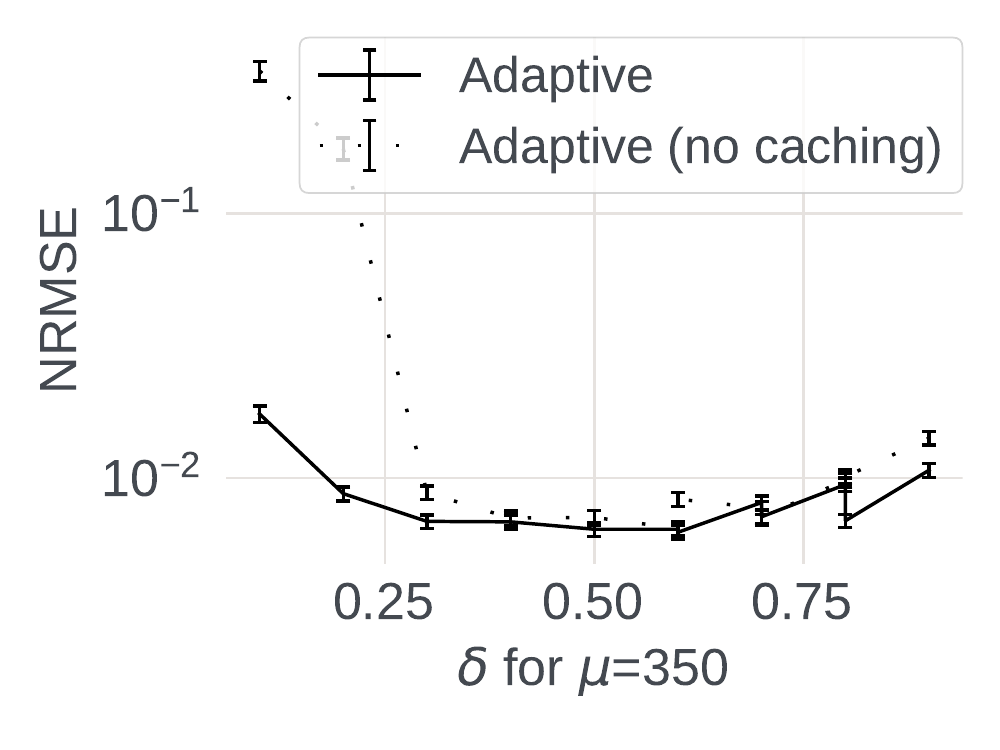}}%
\subcaptionbox{Varying power parameter $\alpha$ \label{fig:bp_alpha}}{\includegraphics[width = 0.33\textwidth]{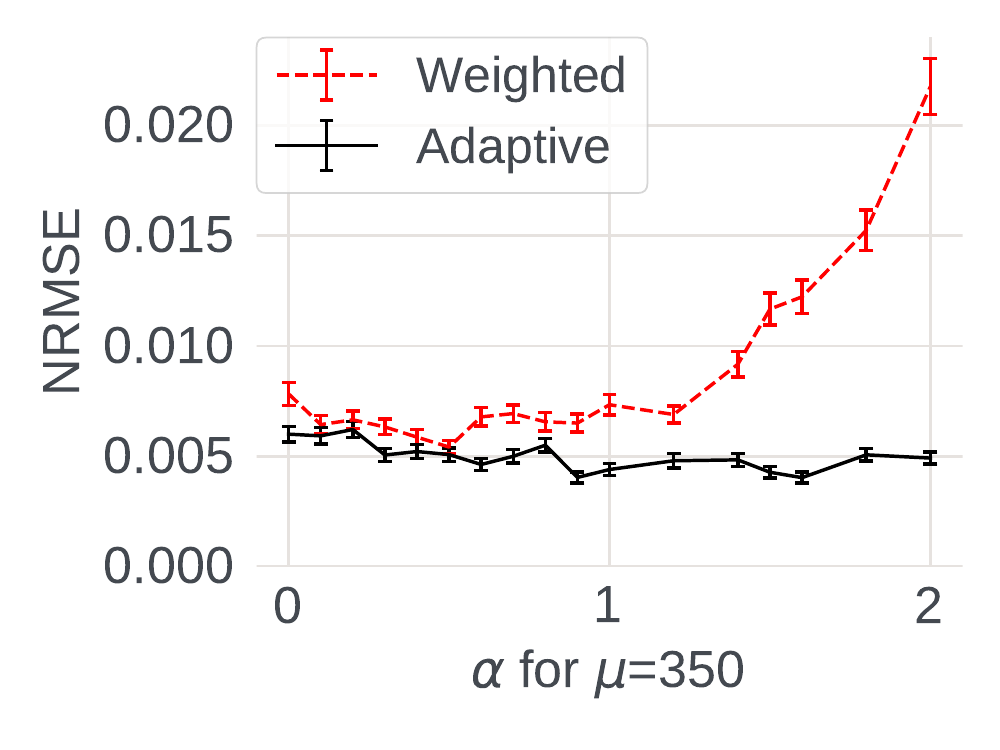}}%
\caption{Parameter Tuning Experiments on Normal data with mean $\mu = 350$ and standard deviation $\sigma = 50$}
\label{fig:param_tuning}
\end{figure*}

\subsection{Parameter Tuning}

We begin with a set of experiments to understand some of the internal parameters in bit pushing and adaptive bit pushing. 
In the first round, 
adaptive bit pushing collects information on all bits of the input, controlled by a parameter $\gamma$, which determines how much weight to put on bit $j$: bit $j$ is sampled in round one with probability $p_j = 2^\gamma$.
Figure~\ref{fig:bp_gamma} shows that the procedure is relatively insensitive to this parameter across the range $0 \leq \gamma \leq 1$, while above 1, the error increases appreciably.  
We do observe that there is a reduction in error for $\gamma = 0.5$, so we adopt this as a default in subsequent experiments. 

The parameter $\delta$ determines how the bit collection is split across the two rounds: 
given a budget of $\clients$ clients, we query $\delta \clients$ clients in round 1 to estimate the bit means, and $(1-\delta)\clients$ in round 2 to build the final estimates. 
Unsurprisingly, setting $\delta$ close to 0 or 1 gives poorer results, and the best results here are found with $\delta$ between $1/3$ and $2/3$. 
This agrees with out theoretical analysis (Section~\ref{sec:adaptivebitpushing}), which predicts that $\delta = 1/3$ optimizes the error. 
We see that adaptive bit pushing with caching (i.e., using the bit means from round 1 as part of the final estimation in round 2) 
is important here, since without caching the accuracy degrades particularly poorly. 
This is because with extreme values of $\delta$, adaptive bit pushing with caching still functions effectively as (single round) weighted bit pushing, where all the information comes from one of the rounds. 

The last parameter we study is $\alpha$, which determines the weights placed on bits in the final round of data collection, so that $p_j \propto 2^{\alpha j}$. 
This applies in both the adaptive and (single round) weighted case. 
In the single round case, there is a clear minimum on this data with $\alpha \sim 0.5$, rather smaller than the theoretical choice of $\alpha=1$.  
In our subsequent experiments, we will show results for both $\alpha = 0.5$ and $\alpha = 1$. 
For the two-round protocol, the dependence on $\alpha$ is much flatter, and choosing $\alpha = 1$ is supported by the data, which we make our default from now on. 
Note that $\alpha = 0$ corresponds to ``uniform weights'': all bits are equally likely to be sampled.  
This rarely performs well in practice, so we do not focus further on this case.
%

The absolute values of NRMSE are also encouraging: the error is typically around 1-2\% of the true mean value. 

\begin{figure*}[t]
\centering
\includegraphics[width=\textwidth]{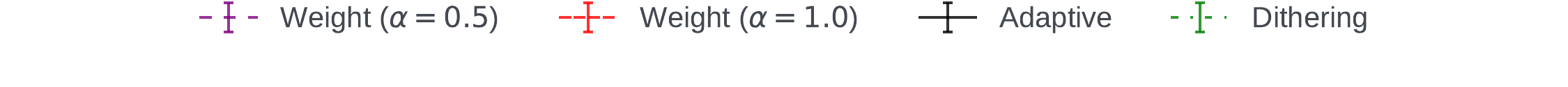}
\subcaptionbox{Estimating mean with $\mu$ varying \label{fig:bp_vary_mu_no_DP}}{\includegraphics[width = 0.33\textwidth]{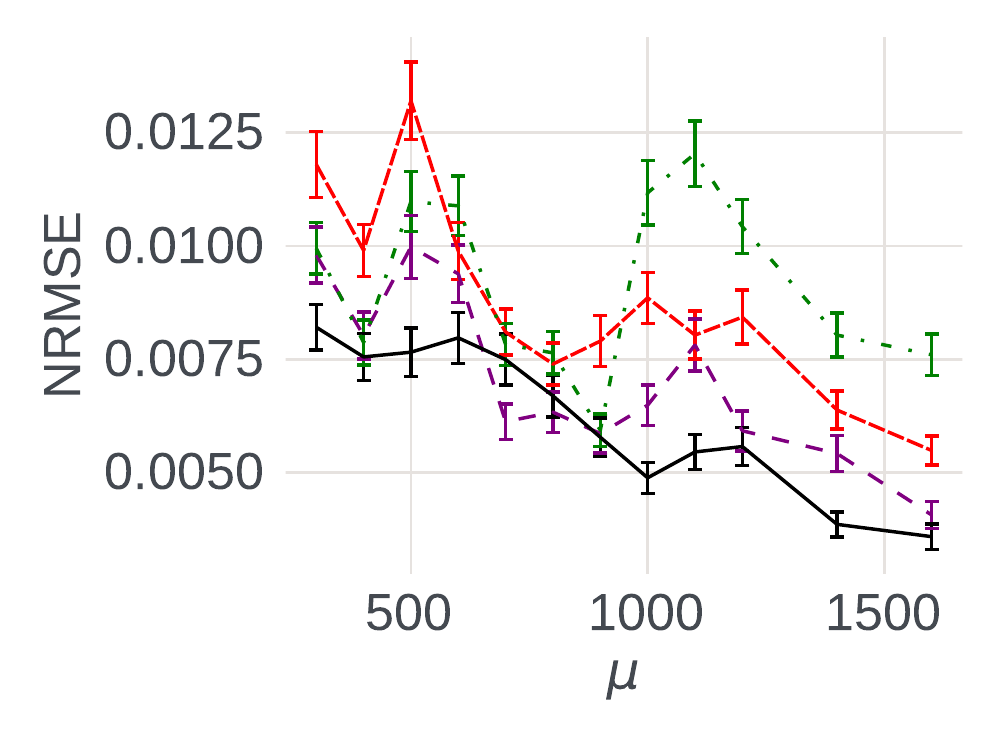}}%
\subcaptionbox{Estimating variance with $\mu$ varying \label{fig:bp_var_no_DP}}{\includegraphics[width = 0.33\textwidth]{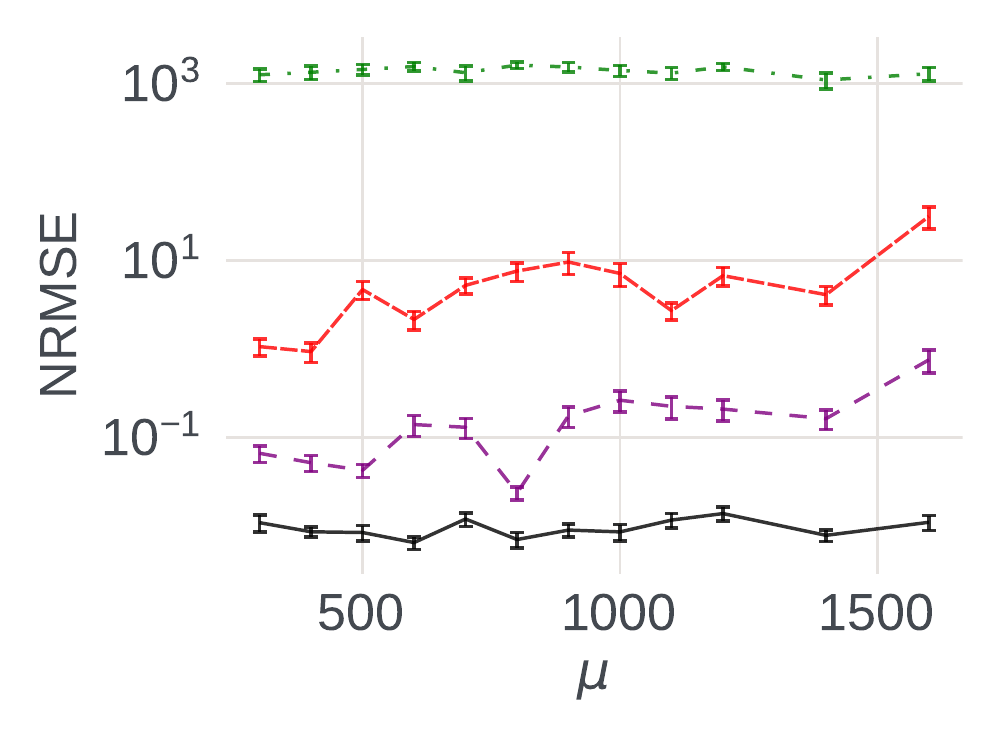}}%
\subcaptionbox{Estimating mean with varying bit depth \label{fig:bp_vary_bits_no_DP}}{\includegraphics[width = 0.33\textwidth]{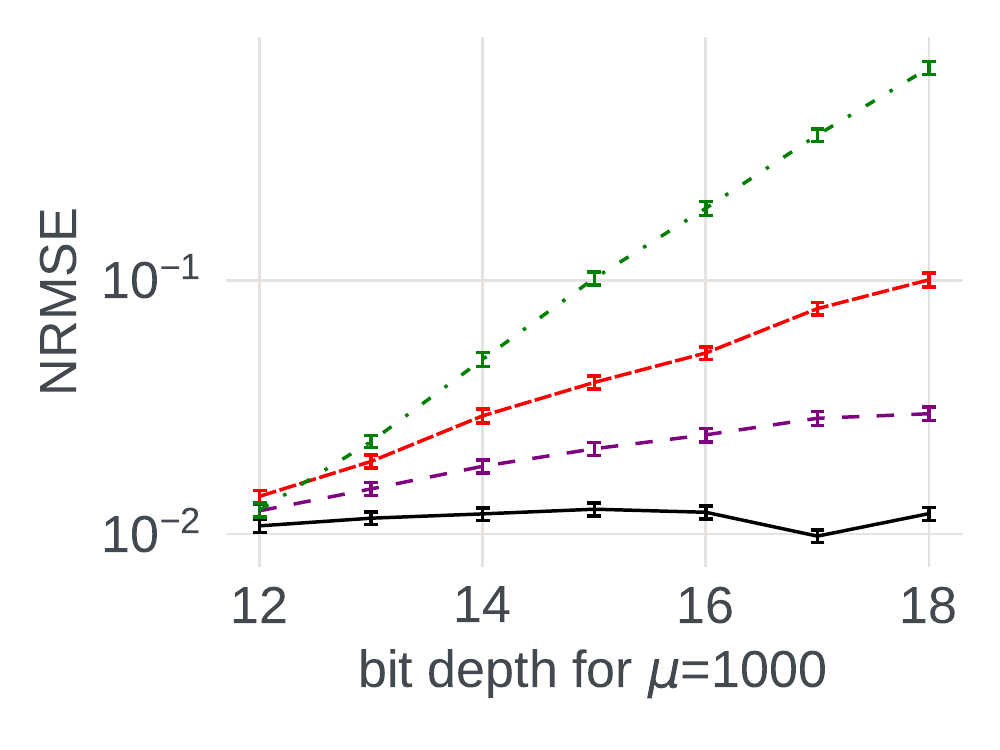}}%
\caption{Accuracy experiments on Normal distributed data with standard deviation $\sigma = 100$}
\label{fig:accuracy}
\end{figure*}

\begin{figure*}[t]
\centering
\includegraphics[width=\textwidth]{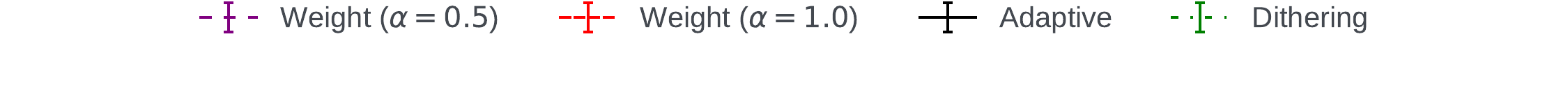}
\subcaptionbox{Estimating mean with varying $\clients$ \label{fig:bp_vary_N_no_DP}}{\includegraphics[width = 0.33\textwidth]{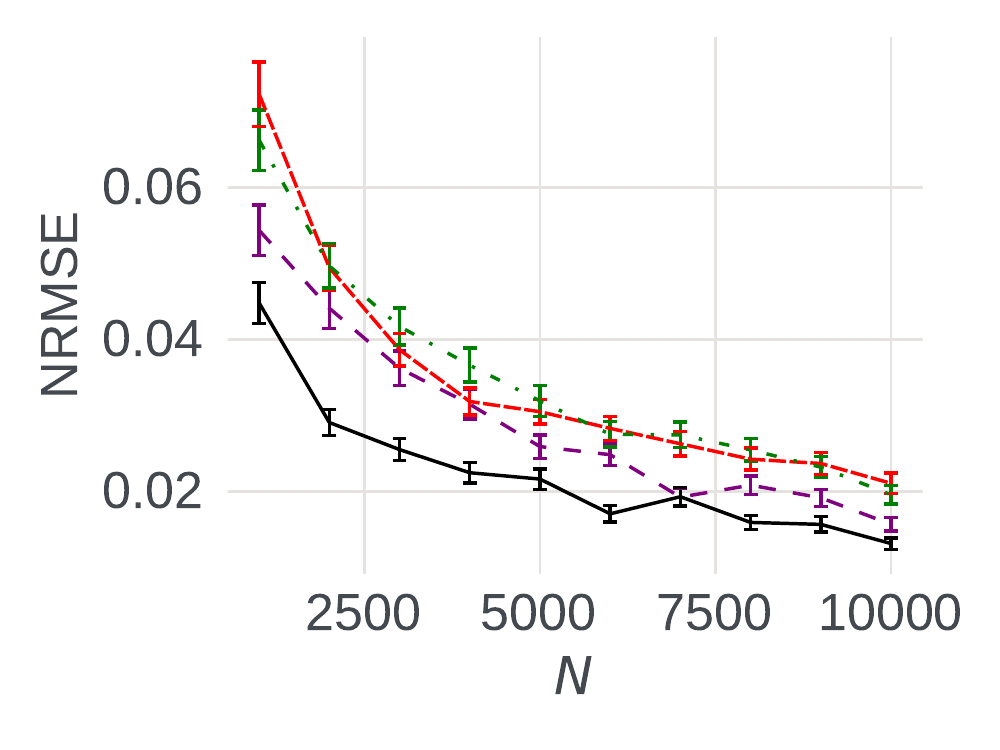}}%
\subcaptionbox{Estimating variance with varying $\clients$ \label{fig:bp_vary_N_var_no_DP}}{\includegraphics[width = 0.33\textwidth]{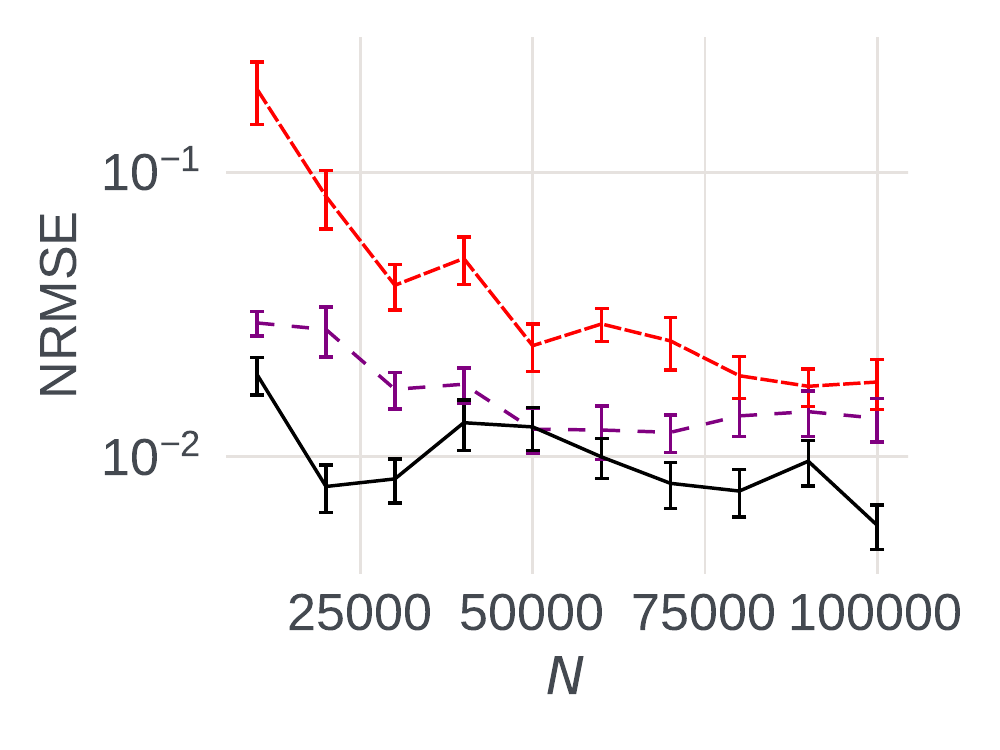}}%
\subcaptionbox{Estimating mean with varying bit depth \label{fig:bp_vary_bits_real_no_DP}}{\includegraphics[width=0.33\textwidth]{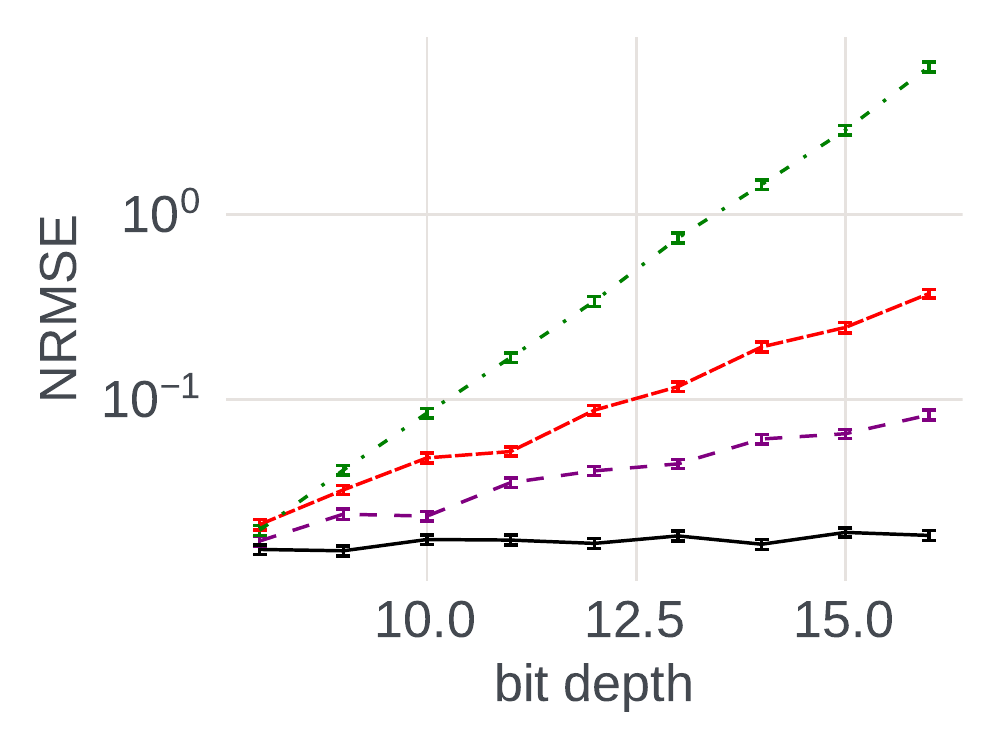}}
\caption{Accuracy experiments on real data}
\label{fig:varyN}
\end{figure*}

\subsection{Accuracy Experiments}

Our second set of experiments focus on accuracy (Figure~\ref{fig:accuracy}).  
Here, we evaluate the {\em subtractive dithering} approach as an alternative approach which reveals only one bit of information about the client's value. We compare against the single round and adaptive bit pushing techniques, with parameters set as determined in the previous section. 

Figure~\ref{fig:bp_vary_mu_no_DP} shows the accuracy as we vary the mean of the input (Normal) distribution. 
There is a general trend for the normalized error to decrease, as the normalizing constant increases faster than the magnitude of the errors. 
This is most notable for the dithering approach around powers of two, where there is a step up in error at the point where we increase the bound on the input values by a factor of 2. 
As in the previous experiments, in a single round, choosing weights based on $\alpha = 0.5$ generally leads to more accurate results. 
Across the whole domain, we find that the adaptive approach reliably achieves the least error. 

We see a different set of behaviors when we attempt to estimate the variance of the distribution, in Figure~\ref{fig:bp_var_no_DP}. 
This is a harder task, as evidenced by error values, which are substantially larger. 
We allocate a larger cohort of 100,000 clients to this task. 
Here, the dithering approach is orders of magnitude worse, due to its inability to adapt to the scale of the input values. 
Among the weighted approaches, it is now the $\alpha = 0.5$ results that are preferred. 
However, the adaptive approach achieves the best accuracy. 
The larger user cohort helps keep the errors manageable: the adaptive approach keeps the normalized errors down in the 1-2\% range. 
Increasing the number of clients $\clients$ would improve the accuracy further. 

We further study the importance of finding accurate bounds on the magnitude of the quantities involved in Figure~\ref{fig:bp_vary_bits_no_DP}. 
We vary the ``bit depth'', which is the number of bits $b$ used in the bit-pushing algorithms (so $2^b$ is the bound used for the dithering approach). 
We see that all the one-round approaches grow in error as $b$ increases: less so for $\alpha = 0.5$, since less weight is apportioned to the (vacuous) high order bits than in the $\alpha = 1.0$ case. 
The adaptive approach, meanwhile, is able to identify the redundant bits in the first round, and discards them in round two. 
Hence, it is largely oblivious to the increase in bit depth. 

A corresponding set of experiments varying $N$ on real data are shown in Figure~\ref{fig:varyN}.
As expected, the normalized error for both mean (Figure~\ref{fig:bp_vary_N_no_DP}) and variance (Figure~\ref{fig:bp_vary_N_var_no_DP}) estimation 
tends to decrease as $\clients$ increases, broadly consistent with the predicted dependence on $\clients^{-1/2}$. 
Unsurprisingly, there is some fluctuation, with the adaptive approach showing more variability for smaller values of $\clients$ for variance estimation. 
Again, we see that the adaptive approach handles the increasing number of bits the best of the methods (Figure~\ref{fig:bp_vary_bits_real_no_DP}).

\subsection{Differential Privacy}
\label{sec:experiment_dp}

\begin{figure*}[t]
\centering
\includegraphics[width=\textwidth]{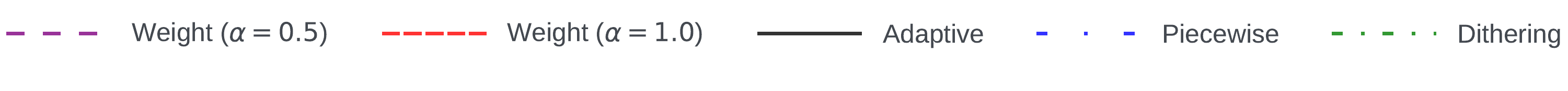}
\subcaptionbox{Varying $\epsilon < 1$ to estimate mean \label{fig:bp_small_epsilon}}{\includegraphics[width = 0.45\textwidth]{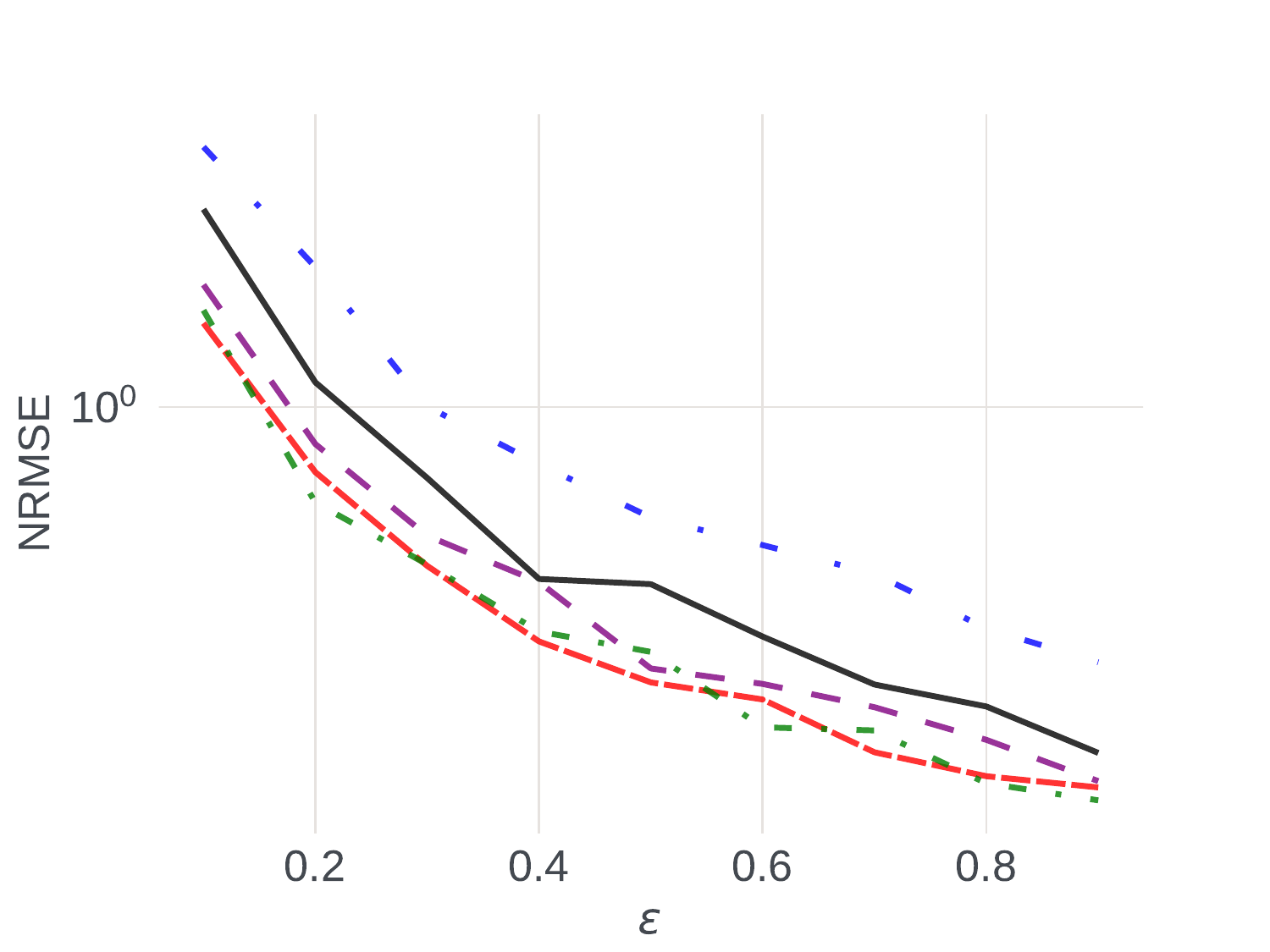}}%
\subcaptionbox{Varying $\epsilon > 1$ to estimate mean \label{fig:bp_big_epsilon}}{\includegraphics[width = 0.45\textwidth]{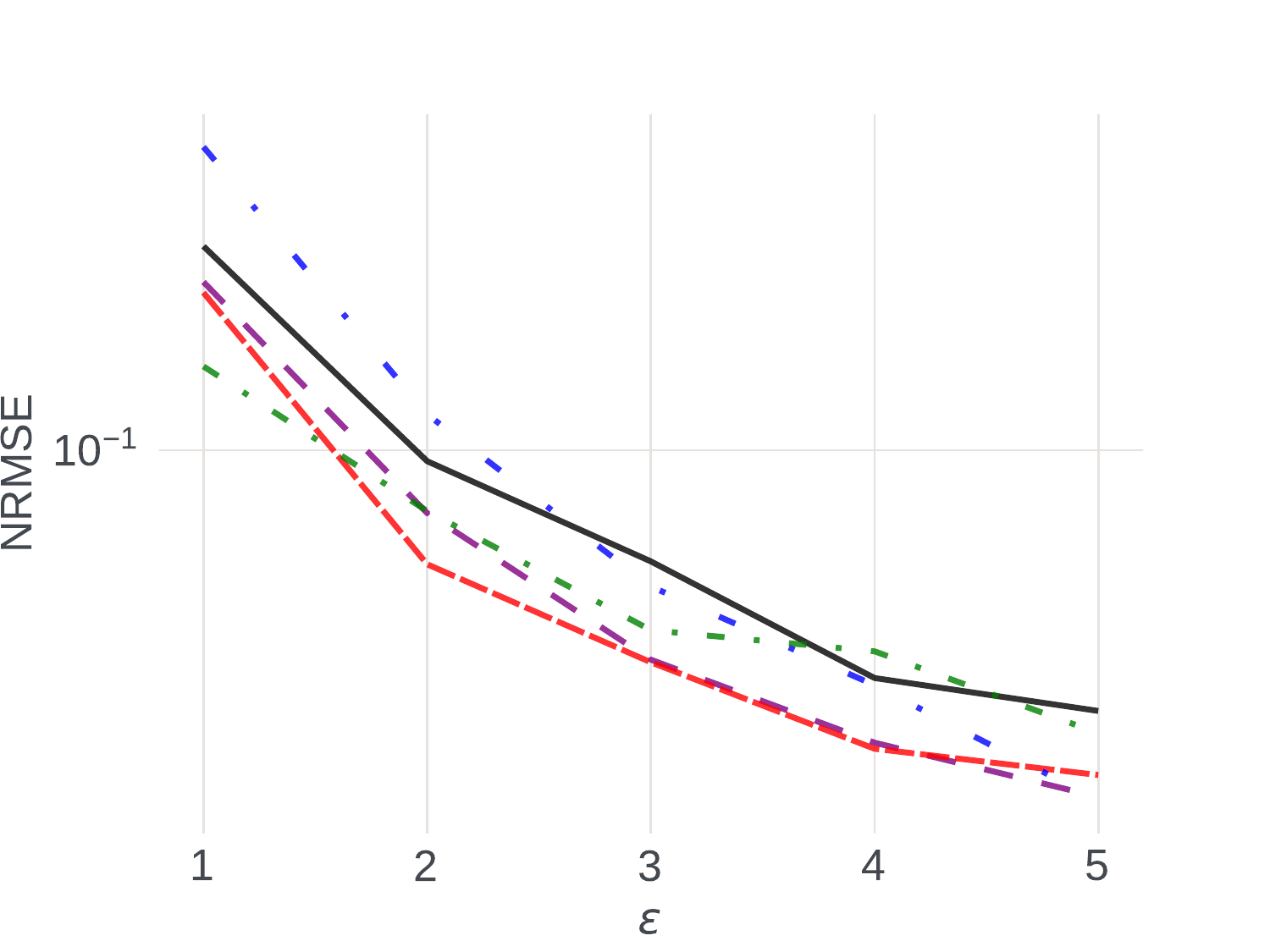}}%
\caption{Accuracy experiments on real data with differential privacy}
\label{fig:varyeps}
\end{figure*}

\begin{figure*}[t]
\centering
\subcaptionbox{Varying the threshold for {\em bit squashing} \label{fig:bp_squashing}}{\includegraphics[width = 0.33\textwidth]{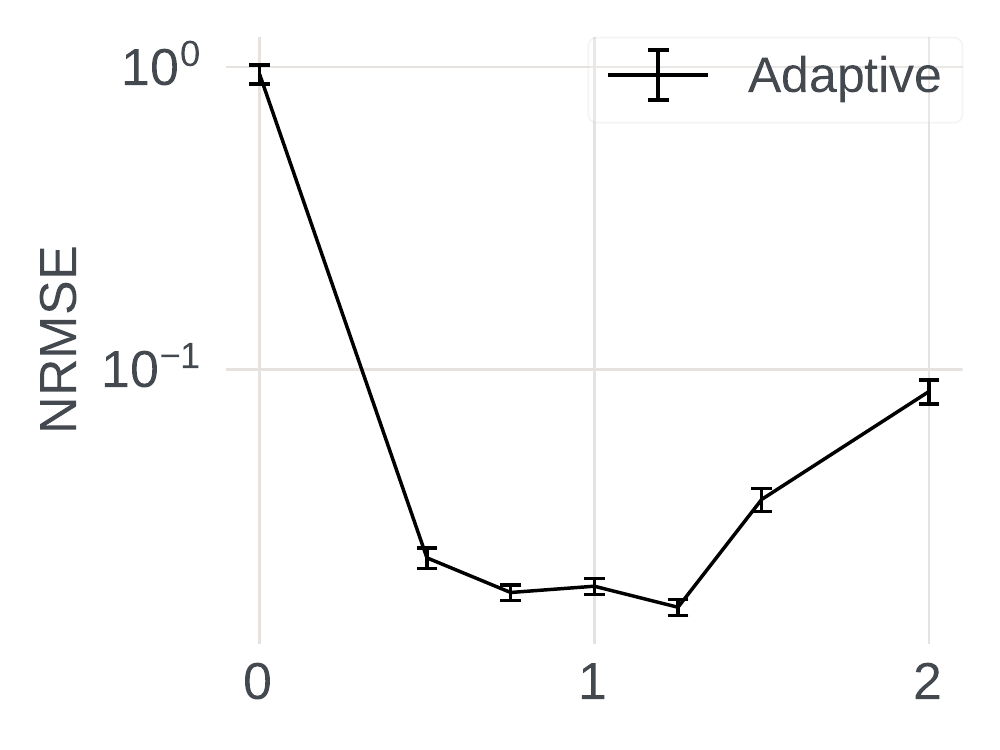}}%
\subcaptionbox{Histogram of bit means\label{fig:bp_DP_histogram}}{\includegraphics[width=0.33\textwidth]{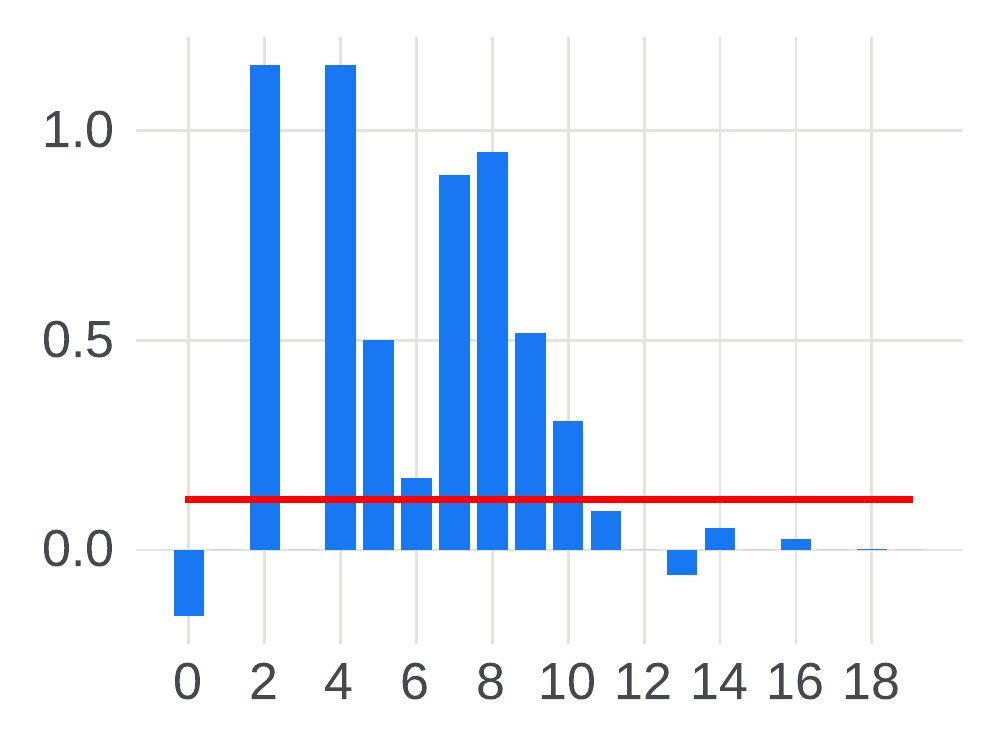}}%
\subcaptionbox{Varying bit depth under DP, $\epsilon=2.0$ \label{fig:bp_vary_bits_DP}}{\includegraphics[width = 0.33\textwidth]{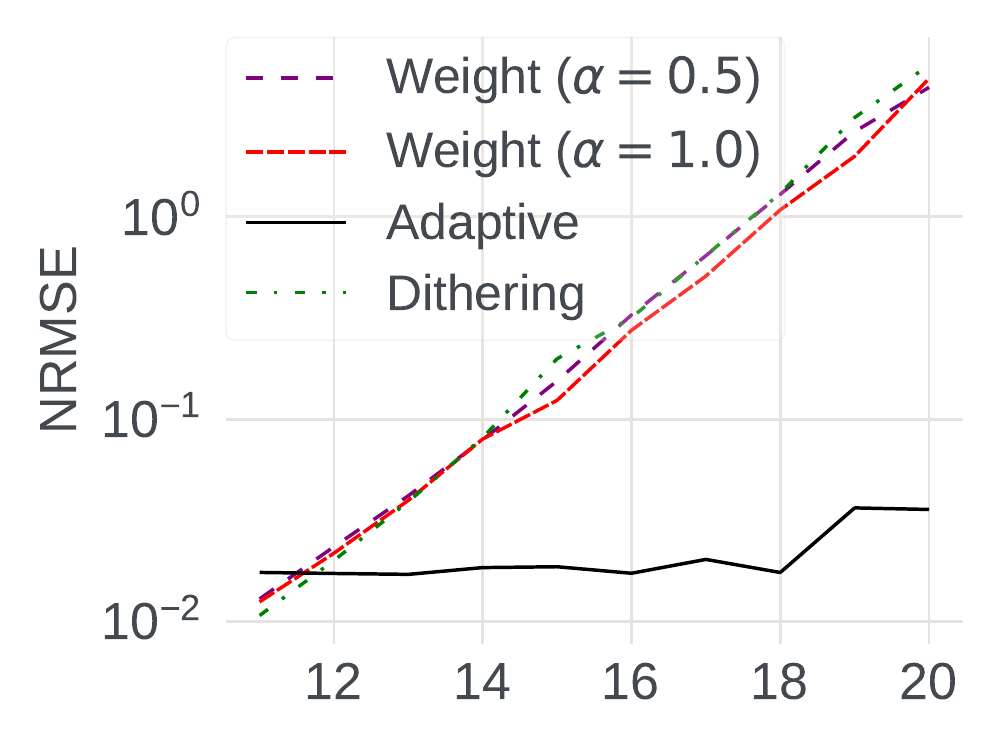}}%
\caption{Accuracy experiments on synthetic data with differential privacy}
\label{fig:dp_accuracy}
\end{figure*}

In this section, we study the impact of providing a differential privacy (DP) guarantee on mean estimation. 
We additionally consider the ``piecewise'' mechanism~\cite{Wang2019}, as well as our previous one-bit methods augmented with randomized response to provide a DP guarantee. 
Figure~\ref{fig:varyeps} shows the RMSE accuracy as we vary the privacy parameter $\epsilon$, split into two regimes: high privacy ($\epsilon < 1$), Figure~\ref{fig:bp_small_epsilon}, and moderate privacy ($\epsilon \ge 1$, Figure~\ref{fig:bp_big_epsilon}). 
Here, we omit results for the Laplace mechanism, where the observed error was considerably higher than others, as expected.
On a log scale plot, the lines are fairly closely clustered, but we see that in this experiment, the single round approach with $\alpha= 1.0$ achieves the least error.  Only when $\epsilon > 3$ do we see points where the adaptive and piecewise approaches achieve lower error. 
Note that the absolute value of RMSE is an order of magnitude larger than without DP noise. 
This is all consistent with our theoretical analysis, where we showed that the variance depends on $\epsilon$ (as $\frac{\exp \epsilon}{(\exp \epsilon - 1)^2}$), and is independent of the value of the bit means.  
Hence, there is no obvious advantage for the adaptive approach, which tries to focus its attention on the bits with higher variance. 

Because of the DP noise, we cannot rely on the bit means of unused bits to be zero.  
Instead, we should apply some filtering to determine which bits are mostly noise, and should have their weight reduced. 
This is captured in Figure~\ref{fig:bp_squashing}, where we apply a simple heuristic: if the value of a bit mean is less than a threshold, we assume that this bit is capturing noise, and `squash' it (i.e., we downweigh its importance). 
The plot shows the effect on RMSE as we vary the threshold, as a multiple of the expected amount of DP noise. 
It turns out that applying a threshold of $1 \pm 0.25$ is very effective at improving accuracy by almost two orders of magnitude. 
Figure~\ref{fig:bp_DP_histogram} shows an example in more detail, with 
a histogram of the estimated bit means for the noisy data with $\epsilon = 2$. 
We see that the DP noise causes some of these estimates to exceed 1.0 or fall below 0.0 (when the DP subtrahend exceeds the true mean).
However, there is a clear ``dense'' region up to bit 11, with higher bits showing random noise. 
The bit-squashing approach treats bits 12 and above as noise, and bases the estimate on bits 0-11 only. 
Figure~\ref{fig:bp_vary_bits_DP} shows this in practice as we increase the bit depth: the adaptive approach using bit squashing maintains the same level of accuracy, while all other methods grow in error proportional to the magnitude of the (noisy) values. 




 



\section{Conclusions and Perspectives}
\label{sec:conclusions}

We have seen that the bit-pushing approach offers strong numerical accuracy with a simple privacy promise: at most one bit of a client's value is disclosed to the server, and such individual bits are shared with plausible deniability via randomized response~\cite{RandResponse1965}.
Our theoretical analysis is confirmed by empirical evaluation across a range of settings. Compared to prior techniques, bit pushing is particularly effective when aggregated values cluster in a narrow range not known in advance.

\paragraph{Notions of optimality} for single-bit estimates have been explored in the recent literature, and methods such as {\em subtractive dithering} were shown optimal~\cite{Ben-Basat2020}. 
Hence, it may look surprising that bit pushing empirically outperforms those methods. 
This apparent paradox stems from the assumptions made in prior proofs of optimality. 
In particular, optimality is invalidated when the true mean can be narrowed down further within a fixed subrange $[0,1]$, and this bracketing is performed by adaptive bit pushing using the same type of inputs used by other protocols. 
In other words, prior optimality results can provide loose bounds in practice, and in some practical cases we improve accuracy by orders of magnitude.

\paragraph{Limitations of the approach} are the flip side of its advantages: when a tight bound on the values \textit{is} known in advance, then bit pushing and existing methods attain similar accuracy. However, accuracy is not the only relevant metric.
\hush{and there is not much to choose between them.} 
In many cases, a single bit of the client's input does not reveal any sensitive information, and so bit pushing alone can provide an intuitive privacy promise to non-experts. When some bits of a value can be privacy-revealing (say, disclosing whether a value is above or below a threshold), plausible deniability for communicated bits is ensured using differential privacy (via randomized response)~\cite{RandResponse1965}. 

\paragraph{Communication costs} of our approach are low, since only a single private bit of data is disclosed. 
However, there are additional overheads to include header information, and list which bit was sampled, so the distinction between sending a single bit versus a few numeric values is not meaningful: both can be easily communicated within a single (encrypted) network packet. 
In settings where each client sends multiple bits, or reveals information about multiple features, the communication benefits become more apparent.

\paragraph{Robustness to poisoning attacks} is a concern for many LDP algorithms~\cite{Cheu2021, Balcer2021}. 
We want to understand how easy it is for a group of clients to collude to force an erroneous estimate. 
Since the estimate is built by averaging over all client reports, no one client can influence the outcome significantly. 
However, if the clients can choose which bit to report, then an adversarial client could choose the most significant bit ($b_{\max}$) 
and deterministically send a 1 (say), to bias the result upwards. 
In this setting, central randomness, where the server picks which bits to report (Section~\ref{sec:bitpushingbasic}), can reduce the impact of a such poisoning attacks. 

\paragraph{Extensions of bit pushing} can be easy to build, but must be analyzed carefully. For example, it is straightforward to apply bit pushing when aggregating gradient updates in Federated Learning. However, gradient compression techniques already provide significant savings in communication, which warrants more comparisons and possibly hybrids~\cite{Vargaftik2021,AppleFed2021}.
It is also straightforward to apply these methods to other statistical estimation tasks. 
For example, covariance estimation is performed by building the Hessian matrix of the input data. 
We can easily apply mean estimation at the client level, by having each client reveal information about the contribution of their input to the Hessian. 
This allows estimation of the correlation coefficient, as well as related quantities such as the Kendall rank correlation coefficient.

\paragraph{Non-numeric data} often arises in Federated Learning applications, i.e., categorical values (such as colors or account types) and sparse features (such as user IDs). In order to support (1) privacy metering, and (2) differential privacy with randomized response, it is natural to represent those values with bits and transmit individual bits. 
However, mean estimation is not a perfect match for such tasks, and when sketching their distribution, bit correlations that we could largely neglect when dealing with numeric data (due to linear separability) can become significant. Future work will seek to extend bit pushing notions to gather correlated bits. 
 
%

\paragraph{Privacy metering at the bit level} can be used in conjunction with bit pushing and differential privacy to provide stronger privacy guarantees and help the general public improve trust in technology. Our work shows how to communicate fewer private bits when aggregating data, and metering private bits shared by an edge device may provide a language more accessible and more convincing to the general public than the language of differential privacy. While counting private bits may give a reasonable metric, one should also consider tallying {\em fractional private bits} when randomized response is used. For example, if a private bit is transmitted with a 50\% chance and otherwise replaced with a non-private (possibly random) bit, this would count as half a bit. However, reasoning about fractional bits may be challenging to the public.

\paragraph{Privacy infrastructure} is needed to turn our proposal into a reality. Our approach draws a red line and limits private bits shared by the device, but the device will naturally share additional (non-private) bits to implement the protocol. 
Platforms for federated analytics and federated learning should provide configurable aggregation services that would package private bits into larger network packets and provide layers of separation to rule out the mixing of private and non-public bits. With such infrastructure, it should be relatively easy to add privacy metering and monitoring, potentially giving the users greater control of their data privacy.

\paragraph{Acknowledgments.}
We are grateful to colleagues who provided input and feedback on this work, including Ilya Mironov, 
Akash Bharadwaj, Kaikai Wang, Harish Srinivas, Peng Chen, and Dzmitry Huba.

\bibliographystyle{plain}
\bibliography{bitpushing}

\begin{thebibliography}{10}

\bibitem{Balcer2021}
Victor Balcer, Albert Cheu, Matthew Joseph, and Jieming Mao.
\newblock Connecting robust shuffle privacy and pan-privacy.
\newblock In D{\'{a}}niel Marx, editor, {\em Proceedings of the 2021 {ACM-SIAM}
  Symposium on Discrete Algorithms, {SODA} 2021, Virtual Conference, January 10
  - 13, 2021}, pages 2384--2403. {SIAM}, 2021.

\bibitem{Ben-Basat2020}
Ran Ben{-}Basat, Michael Mitzenmacher, and Shay Vargaftik.
\newblock How to send a real number using a single bit (and some shared
  randomness).
\newblock {\em CoRR}, abs/2010.02331, 2020.

\bibitem{Cheu2021}
Albert Cheu, Adam~D. Smith, and Jonathan~R. Ullman.
\newblock Manipulation attacks in local differential privacy.
\newblock {\em J. Priv. Confidentiality}, 11(1), 2021.

\bibitem{AppleDP2017}
{Differential Privacy Team at Apple}.
\newblock Learning with privacy at scale.
\newblock
  \url{https://machinelearning.apple.com/research/learning-with-privacy-at-scale},
  December 2017.

\bibitem{Ding2017}
Bolin Ding, Janardhan Kulkarni, and Sergey Yekhanin.
\newblock Collecting telemetry data privately.
\newblock In {\em NeurIPS}, pages 3571--3580, 2017.

\bibitem{DPLimits}
Josep Domingo-Ferrer, David Sánchez, and Alberto Blanco-Justicia.
\newblock The limits of differential privacy (and its misuse in data release
  and machine learning).
\newblock {\em to appear in Comm. ACM}, 2021.

\bibitem{Duchi2018}
J.~C. Duchi, M.~I. Jordan, and M.~J. Wainwright.
\newblock Minimax optimal procedures for locally private estimation.
\newblock {\em Journal of the American Statistical Association},
  113(521):182–201, 2018.

\bibitem{Dwork2014}
Cynthia Dwork and Aaron Roth.
\newblock The algorithmic foundations of differential privacy.
\newblock {\em Foundations and Trends in Theoretical Computer Science}, 9(3-4),
  2014.

\bibitem{Erlingsson2014}
{\'{U}}lfar Erlingsson, Vasyl Pihur, and Aleksandra Korolova.
\newblock {RAPPOR:} randomized aggregatable privacy-preserving ordinal
  response.
\newblock In {\em Proceedings of the 2014 {ACM} {SIGSAC} Conference on Computer
  and Communications Security, Scottsdale, AZ, USA, November 3-7, 2014}, pages
  1054--1067. {ACM}, 2014.

\bibitem{AppleFed2021}
Matthias~Paulik et~al.
\newblock Federated evaluation and tuning for on-device personalization: System
  design \& applications.
\newblock {\tt arXiv:2102.08503}, February 2021.

\bibitem{FacebookIOS14ads}
{Facebook Business Help Center}.
\newblock How {A}pple’s i{OS} 14 release may affect your ads and reporting.
\newblock \url{https://www.facebook.com/business/help/331612538028890}.

\bibitem{Gan2018}
Edward Gan, Jialin Ding, Kai~Sheng Tai, Vatsal Sharan, and Peter Bailis.
\newblock Moment-based quantile sketches for efficient high cardinality
  aggregation queries.
\newblock {\em Proc. {VLDB} Endow.}, 11(11):1647--1660, 2018.

\bibitem{AppleIOS14}
Chaim Gartenberg.
\newblock Why {A}pple’s new privacy feature is such a big deal.
\newblock {\em The Verge}, April 2021.

\bibitem{AppleDPpoor2017}
Andy Greenberg.
\newblock How one of {A}pple's key privacy safeguards falls short.
\newblock {\em Wired}, September 2017.

\bibitem{Haney2020}
Sam Haney, William Sexton, Ashwin Machanavajjhala, Michael Hay, and Gerome
  Miklau.
\newblock Differentially private algorithms for 2020 census detailed dhc race
  \& ethnicity.
\newblock {\em CoRR}, 2020.

\bibitem{IntelSGX}
{Intel Specialized Development Tools}.
\newblock Software guard extensions.
\newblock
  \url{https://software.intel.com/content/www/us/en/develop/topics/software-guard-extensions.html}.

\bibitem{LocalDP2011}
S.~P. Kasiviswanathan, H.~K. Lee, K.~Nissim, S.~Raskhodnikova, and A.~Smith.
\newblock What can we learn privately?
\newblock {\em SIAM Journal on Computing}, 40(3):793–826, 2011.

\bibitem{Levy2020}
Dan Levy.
\newblock Speaking up for small businesses.
\newblock
  \url{https://www.facebook.com/business/news/ios-14-apple-privacy-update-impacts-small-business-ads},
  December 2020.
\newblock Facebook for Business.

\bibitem{Luo2005}
Zhi{-}Quan Luo.
\newblock Universal decentralized estimation in a bandwidth constrained sensor
  network.
\newblock {\em {IEEE} Trans. Inf. Theory}, 51(6):2210--2219, 2005.

\bibitem{GoogleDP2017}
Brendan McMahan and Daniel Ramage.
\newblock Federated learning: Collaborative machine learning without
  centralized training data.
\newblock
  \url{https://ai.googleblog.com/2017/04/federated-learning-collaborative.html},
  April 2017.
\newblock Google AI Blog.

\bibitem{GoogleFA2020}
Daniel Ramage and Stefano Mazzocchi.
\newblock Federated analytics: Collaborative data science without data
  collection.
\newblock
  \url{https://ai.googleblog.com/2020/05/federated-analytics-collaborative-data.html},
  May 2020.
\newblock Google AI Blog.

\bibitem{USCensus2020}
Mike Schneider.
\newblock Groups: Census privacy tool could hurt voting rights goals.
\newblock
  \url{https://apnews.com/article/technology-race-and-ethnicity-data-privacy-voting-rights-census-2020-907d94c8e280b173dc2942feda181348},
  April 2021.
\newblock Associated Press.

\bibitem{Vargaftik2021}
Shay Vargaftik, Ran~Ben Basat, Amit Portnoy, Gal Mendelson, Yaniv Ben{-}Itzhak,
  and Michael Mitzenmacher.
\newblock {DRIVE:} one-bit distributed mean estimation.
\newblock {\em CoRR}, abs/2105.08339, 2021.

\bibitem{Wang2019}
Ning Wang, Xiaokui Xiao, Yin Yang, Jun Zhao, Siu~Cheung Hui, Hyejin Shin,
  Junbum Shin, and Ge~Yu.
\newblock Collecting and analyzing multidimensional data with local
  differential privacy.
\newblock In {\em {IEEE} International Conference on Data Engineering}, pages
  638--649. {IEEE}, 2019.

\bibitem{LocalDP2020}
Teng Wang, Xuefeng Zhang, Jinyu Feng, and Xinyu Yang.
\newblock A comprehensive survey on local differential privacy: Toward data
  statistics and analysis.
\newblock {\tt arXiv:2010.05253}, October 2020.

\bibitem{Wang2017}
Tianhao Wang, Jeremiah Blocki, Ninghui Li, and Somesh Jha.
\newblock Locally differentially private protocols for frequency estimation.
\newblock In {\em {USENIX} Security Symposium}, pages 729--745. {USENIX}
  Association, 2017.

\bibitem{RandResponse1965}
S.~L. Warner.
\newblock Randomized response: A survey technique for eliminating evasive
  answer bias.
\newblock {\em Journal of the American Statistical Association},
  60(309):63–69, 1965.

\end{thebibliography}

\clearpage
\appendix
\section{Omitted Proofs}

\printProofs

\end{document}